\documentclass[11pt,a4paper]{article}
\usepackage[T1]{fontenc}
\usepackage[utf8]{inputenc}
\usepackage{lmodern,amsmath,amssymb,amsthm,microtype}
\usepackage[a4paper,top=22mm,bottom=22mm,left=24mm,right=24mm,headheight=14pt,headsep=7mm,footskip=11mm]{geometry}
\usepackage{array,booktabs,xcolor,fancyhdr,needspace,float}
\definecolor{ink}{RGB}{29,62,77}
\usepackage[unicode,colorlinks=true,linkcolor=ink,citecolor=ink,urlcolor=ink,bookmarksnumbered=true]{hyperref}
\hypersetup{pdftitle={Short Matrix Refreshes for Leakage-Resilient Arithmetic Circuits},pdfauthor={Tomasz Kazana},pdflang={en}}
\fancypagestyle{firstpage}{\fancyhf{}\fancyfoot[C]{\small\thepage}}
\newtheorem{theorem}{Theorem}[section]
\newtheorem{lemma}[theorem]{Lemma}
\newtheorem{corollary}[theorem]{Corollary}
\theoremstyle{definition}\newtheorem{definition}[theorem]{Definition}
\newcommand{\F}{\mathbb F}
\newcommand{\E}{\mathbb E}
\newcommand{\Prb}{\mathbb P}
\newcommand{\R}{\mathcal R}
\newcommand{\GL}{\mathrm{GL}}
\newcommand{\TV}{d_{\mathrm{TV}}}
\newcommand{\ind}{\mathbf 1}
\newcommand{\orb}[2]{\mathcal O^{(#1)}_{#2}}
\newcommand{\law}[2]{\mu^{(#1)}_{#2}}
\newcommand{\op}[1]{\lVert #1\rVert_{\mathrm{op}}}
\newcommand{\hs}[1]{\lVert #1\rVert_2}
\newcommand{\siv}[1]{\lVert #1\rVert_{S_4}}
\newcommand{\nuc}[1]{\lVert #1\rVert_*}
\newcommand{\ellone}[1]{\lVert #1\rVert_1}
\newcommand{\supn}[1]{\lVert #1\rVert_\infty}
\newcommand{\cyc}{\operatorname{cyc}}
\newcommand{\rank}{\operatorname{rank}}
\newcommand{\tr}{\operatorname{tr}}
\newcommand{\bitlen}{\mathsf L}
\newcommand{\enL}{\mathcal E_L}
\newcommand{\enR}{\mathcal E_R}
\newcommand{\topic}[1]{\par\medskip\noindent\textbf{#1.}\ }
\newcommand{\field}{\F_q}
\newcommand{\partitle}[1]{\par\smallskip\noindent\emph{#1.}\ }
\newcommand{\proofanchor}[1]{\phantomsection\label{#1}}
\newcommand{\proofref}[2]{\hyperref[#2]{Section~\ref*{#1} (p.~\pageref*{#2})}}
\newcommand{\proofnote}[1]{\par\nopagebreak[4]\smallskip\noindent{\normalfont\small #1\par}}
\begin{document}
\begin{center}
{\LARGE\bfseries Short Matrix Refreshes for\\[2pt] Leakage-Resilient Arithmetic Circuits\par}
\vspace{.65em}
{\large Tomasz Kazana\par}
\smallskip
{\small Preprint, 29 November 2026}
\end{center}
\thispagestyle{firstpage}
\begin{abstract}
We study the longstanding problem of \emph{two-component leakage-resilient computation}.
In this setting, two processors repeatedly evaluate a circuit on shared secrets.
Even if each operation leaks a bounded number of bits from each processor's active
memory, the accumulated leakage should reveal essentially no more about the secrets
than the public inputs and outputs already do.

Dziembowski and Faust (2012) gave an information-theoretic compiler that transforms
an arithmetic circuit into a functionally equivalent leakage-resilient circuit.
The two processors running the compiled circuit use a leak-free source of correlated
randomness. Goldwasser and Rothblum (2012) proposed a compiler that dispenses with
leak-free hardware by using many components, and Dachman-Soled, Liu and Zhou (2015)
obtained a two-component compiler under computational assumptions. More recently,
Dziembowski and Faust (2026) showed that matrix refreshing suffices for
information-theoretic security with two processors and no leak-free source.
Their binary-field construction gives the strongest previous leakage bounds we
know for this approach. They left open the extension of their quantitative
guarantees to larger fields.

We resolve this question by giving an information-theoretic compiler for stateful
arithmetic circuits over every finite field. Compared with the binary construction
of Dziembowski and Faust, both our storage protocol and our circuit compiler
tolerate more leakage from each processor, measured as a fraction of one share.
We also reduce communication. For shares of dimension \(n\), a storage refresh
communicates \(O(n)\) field elements and a multiplication communicates \(O(n^2)\).
\end{abstract}
\section{Introduction}\label{sec:intro}

Consider a device that authenticates many messages with the same secret key. On every use an observer sees the message and its tag and can also learn a bounded amount about each of the device's two processors. Can we implement the device so that all these observations together reveal essentially no more about the key than the messages and tags alone?

The difficulty is already present in storage. If the two processors keep the same shares of a secret, an observer can ask for different coordinates on successive uses and eventually recover both shares. The processors can renew the shares between uses, but the renewal itself may leak. For computation there is a further difficulty. The shares produced by one gate become inputs to later gates, so leakage from several operations can reveal more than any one of their answers.

\topic{From leakage-resilient storage to circuits}
The study of leakage-resilient computation grew out of work on wire probing and on what physical observations reveal about an active computation \cite{ISW03,MR04}. Early compilers for continual leakage used protected hardware \cite{JV10,GR10}. A starting point for the present line of work is the inner-product encoding of Dav\`i, Dziembowski and Venturi \cite{DDV10}. A field element \(s\) is represented by two vectors \(L,R\) with \(L^\top R=s\), one at each processor. Dziembowski and Faust showed how to refresh this encoding with a protected source of correlated randomness \cite{DF11} and used such a source to compile arithmetic circuits \cite{DF12}. Andrychowicz gave a linear-work refresh with a protected source \cite{And12}. Dziembowski and Faust give a broader account of the early models in \cite{DF12}.

Goldwasser and Rothblum proposed an inner-product compiler without protected hardware in a model where only the active computation leaks \cite{GR12}. Dziembowski and Faust discuss a gap in the published Goldwasser--Rothblum proof in \cite[App.~C]{DF26}. Later work explored the use of auxiliary computing components \cite{BDL14}, and Dachman-Soled, Liu and Zhou obtained a two-component compiler under computational assumptions \cite{DLZ15}. In their model, leakage may depend on a component's full evaluation state. Here each query reads only one processor's local view of the active operation.

More recently, Dziembowski and Faust \cite{DF26} proved that a uniform matrix refresh protects a stored bit under continual local leakage, and they used it to compile stateful Boolean circuits built from NAND gates. Their two processors need no leak-free source inside the computation. To see what their storage theorem establishes, take nonzero shares \(L,R\), sample a uniform invertible matrix \(M\), and refresh by
\[
                  (L,R)\quad\longmapsto\quad(ML,M^{-\top}R).
\]
The inner product stays equal to \(s\). For any fixed starting encoding of \(s\), the new pair is uniform over the encodings of \(s\) before the round's leakage is observed. The security question arises because both processors know \(M\). A leakage query may inspect the entire matrix description together with the processor's old share, and its answer can determine which question the adversary asks next.

\topic{Our results and comparison with Dziembowski and Faust}
We replace the uniform matrix by one sampled from an explicit family with short descriptions. A storage refresh then sends only \(O(n)\) field elements for shares of dimension \(n\). Our proof gives the same leakage rate for this family as for a uniform matrix (Theorems~\ref{thm:storage} and~\ref{thm:short-storage}). To multiply, each processor first forms the tensor product of its two input shares, as in \cite{DF12}. The processors then refresh the resulting long shares and shorten them back to dimension \(n\). Each tensor share has length \(n^2\), so transmitting a full change of basis for it would cost order \(n^4\) field elements. Our entire multiplication gate communicates \(O(n^2)\) field elements.

Theorem~\ref{thm:circuit} gives statistical security over every finite field \(\field\), uniformly even when \(q\) grows with \(n\). Dziembowski and Faust prove their quantitative bounds over \(\mathbb F_2\) and leave the extension to larger fields open in \href{https://arxiv.org/html/2611.04876v1\#S6.SS4}{Section~6.4} of their paper (\textit{General fields}, p.~75). Here a gate computes \(u+v\), \(uv\), or \(c-uv\) directly in \(\field\), for public \(c\). For shares in \(\field^n\), we normalize per-processor leakage by \(\bitlen=n\log_2 q\). At any fixed per-operation rate below \(1/230\), a simulator given only the public inputs and authorized outputs reproduces the leakage across repeated evaluations with statistical error \(2^{-\Omega(\bitlen)}\), provided the total number of operations is polynomial in \(\bitlen\) and \(n\) tends to infinity. For example, over \(\mathbb F_{2^\ell}\) a wire carries \(\ell\) bits and the leakage allowance scales with \(n\ell\). Table~\ref{tab:rates} compares the rates under the same per-party budgets.

\begin{table}[H]
\centering\small
\begin{tabular}{@{}lcc@{}}
\toprule
Quantity & Dziembowski--Faust~\cite{DF26} & This work\\
\midrule
Field covered by the security theorems
  & \(\mathbb F_2\) & Every finite field\\
\addlinespace
Storage leakage per party per round
  & \(<1/54\) & \(<1/32\)\\
Circuit leakage per party per operation
  & \(<1/486\) & \(<1/230\)\\
Circuit leakage per party per phase
  & \(<1/972\) & \(<1/305\)\\
\addlinespace
Bits sent per storage refresh
  & \(n^2\) & \(14\lceil n/2\rceil\)\\
Bits sent per NAND gate
  & \(\Theta(n^4)\) in expectation & \(\Theta(n^2)\)\\
Expected bit operations per storage refresh
  & \(O(n^3)\) & \(O(n^2)\)\\
Expected bit operations per NAND gate
  & \(O(n^6)\) & \(O(n^4)\)\\
\bottomrule
\end{tabular}
\caption{Fields, sufficient leakage rates, and implementation costs.
Per-party leakage budgets in bits are normalized by \(n\) in~\cite{DF26}
and by \(n\log_2 q\) here. Each fixed rate strictly below the indicated bound gives
exponentially small statistical error over polynomially many rounds or
operations as \(n\) grows. The per-operation budget covers both phases of
a gate. The per-phase row assumes the same allowance \(b\) per party in
every phase. Setting the total gate budget to \(\lambda=2b\) in the bound
of~\cite{DF26} gives the entry \(1/972\). The four cost rows compare the
implementations over \(\mathbb F_2\).}
\label{tab:rates}
\end{table}

For a per-party budget of \(\lambda\) bits, the statistical error bounds in \cite{DF26} are \(O(T\,2^{6\lambda-n/9})\) for \(T\) storage rounds and \(O(S\,2^{18\lambda-n/27})\) for \(S\) circuit operations. Our corresponding bounds are \eqref{eq:storage-bound} and \eqref{eq:circuit-main}. If each phase receives its own budget, our stateful compiler tolerates any fixed rate below \(1/305\) per party per phase. For a single gate on independently encoded inputs with its output shares hidden, Corollary~\ref{thm:gate} gives rates below \(1/9\) per gate or \(1/18\) per phase. The circuit theorem also accounts for what later gates reveal when they reuse those shares.

Over general \(\field\), our communication costs have the same orders when counted in field elements (plus one flag bit per gate), and work is counted in base-field operations. The \hyperref[proof:implementation]{operation counts in Section~\ref*{pr:gate}} give the last two rows of Table~\ref{tab:rates} for our protocols. The binary protocols of \cite{DF26} apply for \(n\ge2\). Our short matrix descriptions require \(n\ge8\) for storage and \(n\ge9\) for product gates.

In both compilers the choice of leakage functions in later phases may depend on answers from earlier phases. Our analysis also permits each query within a phase to depend on earlier answers from that phase, as sketched in Section~6.2 of \cite{DF26}. Dziembowski and Faust formalize their main binary results in Lean. In the earlier protocol game of \cite[Sec.~2.1]{DF12}, the leakage can also read accumulated local coins and messages after a protocol. Here the erasures specified at phase boundaries determine which data remain available.

\topic{How the argument fits together}
For storage, we follow one possible leakage history at a time. Its probability is the average of a table over all encodings of the secret. Truncating the singular values of these tables isolates the part that can survive another refresh. We first explain the argument for a uniform matrix. Only then do we identify the few images of \(M\) that the proof actually examines and construct a much shorter random description with the required properties.

For a gate, the relevant table answers a different question. Once the input and output encodings are specified, how likely is a given answer from this operation? We show that this table is well approximated by a sum of products of left-only and right-only functions. These local approximations can be combined even at wires shared by several operations. The proofs are in Section~\ref{sec:proofs}. Appendix~\ref{sec:tools} gives a self-contained account of the spectral estimates in the normalization used here.

\topic{Project history and relation to \cite{DF12,DF26}}
Stefan Dziembowski and Sebastian Faust used a protected source of correlated randomness in their compiler \cite{DF12}. After its publication, I worked with Stefan Dziembowski, then my doctoral supervisor, on refreshing inner-product shares without that source. I proposed the matrix refresh protocol during that collaboration. We could not prove its security under continual leakage and eventually set the project aside.

Years later we returned to the problem separately, neither side aware that the other had resumed work. I was writing this manuscript when I learned of their paper \cite{DF26}. By then I had already obtained a proof of security under continual leakage independently of \cite{DF26} and had begun extending the result to circuits. Thus, the circuit construction and its security proof draw on both my earlier ideas and techniques from their paper. Specifically, the compiler follows their schedule for refreshing gate inputs, and in the security proof I adapt their counting argument for matrices with prescribed row and column images and their expansion over marked operations and wires.

\topic{AI disclosure}
I feel that while I provided the overarching idea -- namely, the matrix refresh protocol and the proposal to investigate whether the distribution of the shares, conditioned on the leakage observed so far, could be approximated by a weighted sum of indicators of large rectangles -- ChatGPT essentially produced the substance of the paper, including the key technical ideas and the hardest proofs. Specifically, it proposed both an explicit family of refresh matrices with short descriptions and the use of truncated singular value decompositions to analyze the tables of transcript probabilities, built the mathematical framework, derived the bounds on accumulated truncation error, performed the calculations, and worked out the detailed proofs. It also assisted in drafting the manuscript and writing code to verify individual cases. I edited the text in the hope of making the reasoning clear and intuitive for a human mathematician. I take full responsibility for the correctness of the final version of the paper; however, readers should treat this preprint as provisional, as subsequent versions may include corrections.

\section{Encodings and leakage}\label{sec:model}

The two processors, called the left and right parties, hold shares that change at every refresh. The adversary keeps a record of its leakage answers across all rounds. Storage security asks how much the distribution of this record can depend on the secret.

\topic{\proofanchor{claim:encoding}Encoding one field element}
Fix a finite field \(\field\) of prime-power order \(q\) and an ordinary share dimension \(n\ge4\). To encode \(s\in\field\), draw a nonzero left vector \(x\in\field^n\) uniformly. Then draw the right vector \(y\) uniformly from the nonzero solutions of \(x^\top y=s\). The parties store one vector each. Decoding takes their inner product.

We will need the same encoding in other dimensions. For \(d\ge2\), let \(\Omega_d=\field^d\setminus\{0\}\), of size \(N_d=q^d-1\). Write
\[
            \orb d s=\{(x,y)\in\Omega_d\times\Omega_d:x^\top y=s\}
\]
for the set of encodings and \(\law d s\) for the uniform distribution on it. The sampling rule above gives this distribution in dimension \(n\), since every nonzero \(x\) has the same number of eligible partners \(y\). By symmetry, either share on its own is uniform for every secret, including zero.

For comparison, let \(\nu_d\) be the law of two independent uniform draws \(x,y\) from \(\Omega_d\). Conditioning this pair on \(x^\top y=s\) gives \(\law d s\), whose density relative to \(\nu_d\) is written explicitly in \eqref{eq:orbit-density}. For ordinary shares we abbreviate \(\mu_s=\law n s\), \(\nu=\nu_n\), and \(N=N_n\).

\topic{One storage round}
Start with \((X_0,Y_0)\sim\mu_s\). In round \(i+1\), let \(\xi\) be the full random tape used by the left party to sample the refresh matrix, including any rejected samples. The left party privately sends the right party a description \(d(\xi)\) that determines the invertible matrix \(M\). The parties compute
\[
       X_{i+1}=MX_i,\qquad Y_{i+1}=M^{-\top}Y_i,
       \qquad M^{-\top}=(M^{-1})^\top .
\]
The communication cost of the round is the length of \(d(\xi)\). The inner product is still \(s\), and only the two new shares survive into the next round.

\topic{What a query can read}
A storage round is one leakage phase. Write \(V_L,V_R\) for the parties' local views during a phase. Each contains that party's active input and output shares, intermediate values, sampling tape (including rejected samples), and messages sent or received. In a storage round the left view contains the full tape \(\xi\), and both views contain the transmitted description \(d(\xi)\). At the phase boundary the parties erase everything in the active workspace except what the protocol retains. Fresh coins are independent of earlier states and answers.

Given all public answers so far, the adversary may end its queries for this phase or choose a party, any function of its view, and an answer length. Let \(f_i\) and \(g_i\) denote the \(i\)-th functions queried on the left and right, respectively, within this phase. For example, on old shares \(x,y\), it can first ask whether \(x_1=0\). After seeing the bit \(u=f_1(V_L)\), it can choose \(g_1\) to test whether \(y_{1+u}=0\). Every path obeys the bit budgets \(\sum_i \lvert f_i(V_L)\rvert\le b_L\) and \(\sum_i \lvert g_i(V_R)\rvert\le b_R\), summing over the queries made on each side. Queries only observe the computation. We fix the adversary's independent randomness and auxiliary information and average over them at the end.

\topic{Transcripts and security}
For a fixed strategy let \(P_s^T\) be the law of the complete leakage transcript after \(T\) storage rounds starting from secret \(s\). To compare two secrets \(s,t\), run the same strategy in each experiment. The total variation distance \(\TV(P_s^T,P_t^T)\) is the largest absolute difference between the probabilities the two laws assign to a transcript event. The expectation bounds for tests of two such laws are given in Appendix~\ref{app:incidence}, \eqref{eq:tv-test}.

\section{Why uniform matrix refreshing is secure}\label{sec:storage}

To make the storage argument easier to follow, we begin with a matrix \(M\) sampled uniformly from \(\GL_n(\field)\). In this case the left party sends the entire matrix. From any fixed encoding of \(s\), one refresh gives a uniform encoding of \(s\) by the one-pair case of Lemma~\ref{lem:orbit}. This is exact freshness \emph{before observing the leakage from this round}. To see the qualification in action, take \(q=2\) and let the left party leak the first coordinate of \(ML\). That single answer fixes the first coordinate of the new left share, precisely the share with which the next round starts. A proof for many rounds must keep track of the dependence between the current shares and the accumulated answers.

\begin{theorem}[Storage with a uniform matrix]\label{thm:storage}
Let \(b_L,b_R\) bound the number of bits leaked on the two sides in each round, and put \(K=2^{b_L+b_R}\). For every \(n\ge4\), every \(s,t\in\field\), and \(T\) rounds of uniform matrix refreshing,
\[
                 \TV(P_s^T,P_t^T)\le CTK^2q^{\,2-n/8}
\]
for an absolute constant \(C\). Thus equal per-party budgets up to \(\rho n\log_2q\), for fixed \(\rho<1/32\), give exponentially small error for polynomially many rounds as \(n\to\infty\).
\proofnote{The proof is given in \proofref{pr:storage}{proof:storage}.}
\end{theorem}

For each possible leakage history we will compare its probability for a fixed secret with its probability when the two initial shares are independent. Both probabilities are averages of the same table. We keep its large singular values and use one more refresh to control the error. The factor \(T\) comes from adding one such error per round.

\topic{One table for each history}
Run one experiment starting from independent uniform shares, leaving their inner product unspecified. Write \(Z_i\) for the transcript after \(i\) rounds. Without conditioning on \(Z_i\), the pair \((X_i,Y_i)\) still has law \(\nu\), since each matrix acts by a separate permutation on each side. For a fixed possible history \(h\) of \(i\) rounds, define the table
\[
             p_h(x,y)=\Prb[Z_i=h\mid X_i=x,\ Y_i=y].
\]
For example, one row fixes the present left share and runs over all choices of the present right share. Its entries answer the same question about the same history \(h\). Conditioning the product experiment on its invariant inner product gives
\begin{equation}
             P_s^i(h)=\Prb_s[Z_i=h]
                      =\E_{(x,y)\sim\mu_s}p_h(x,y).
             \label{eq:history}
\end{equation}
In particular, all secrets use the \emph{same} table. Its average over independent shares gives the reference probability for that history.

How does the table change when the next round produces a transcript \(j\)? Fix its new shares \((x,y)\) and the full sampling tape \(\xi\). The preceding shares must then have been \(a=M^{-1}x,\ b=M^\top y\). \proofanchor{claim:rectangles}For fixed \(h\) and \(\xi\), a complete
transcript \(j\) tells us which questions were asked, even when they
were chosen adaptively. The left answers restrict \(a\) to a set
\(A_j(\xi)\), and the right answers restrict \(b\) to a set
\(B_j(\xi)\). Thus the pairs giving \(j\) form the rectangle
\(A_j(\xi)\times B_j(\xi)\). As \(j\) varies, these rectangles
partition the pairs of old shares. Looking backwards by one round gives \(p_{hj}=\R_jp_h\), where for any table \(H\)
\begin{equation}
       (\R_jH)(x,y)=\E_\xi\big[H(a,b)\,\ind_{A_j(\xi)}(a)\,\ind_{B_j(\xi)}(b)\big],
       \qquad (a,b)=(M^{-1}x,M^\top y).
       \label{eq:round}
\end{equation}
Here \(\ind_A\) denotes the indicator of \(A\), and \(hj\) is the history obtained by appending \(j\) to \(h\). We write \(\R_{h,j}\) when the incoming history matters. For every fixed tape the update preserves \(\nu\), so the endpoint in the product experiment is independent of \(\xi\). Thus the expectation in \eqref{eq:round} uses the original law of the full sampling tape.

\topic{\proofanchor{claim:history-trunc}What to keep from a table}
Let \(w_h=\max_{x,y}p_h(x,y)\) and \(m_h=\E_\nu p_h\). The latter is the probability of \(h\) in the product experiment, so \(\sum_{|h|=i}m_h=1\). To bound the sum \(W_i:=\sum_{|h|=i}w_h\), fix a history \(h\) of length \(i\). For a uniform matrix, the predecessor of a fixed endpoint is a uniform encoding of its inner product. Thus \(w_{hj}\le\max_{s\in\field}P_s^i(h)\) for each extension \(hj\). Reading the answers bit by bit gives a binary tree of depth at most
\(b_L+b_R\), so the history \(h\) has at most \(K=2^{b_L+b_R}\)
possible extensions in this round. Summing over histories and using
\(\sum_{|h|=i}P_s^i(h)=1\) for every \(s\) gives
\[
 W_{i+1}=\sum_{|h|=i}\sum_j w_{hj}
      \le K\sum_{s\in\field}\sum_{|h|=i}P_s^i(h)=Kq .
\]
Initially \(W_0=1\). The number of possible histories can increase at every round while this bound stays fixed. The corresponding recursion for a short sampler is \eqref{eq:history-weights}.

At each \(h\), keep the singular values of \(p_h\) of size at least \(\theta w_h\), with \(\theta=q^{-n/4}\). Call the resulting table \(a_h\), taking \(a_h=0\) for a zero table. The devices never perform this step. It is a way to compare histories of very different probabilities at the appropriate scale. The omitted part has small operator norm by \eqref{eq:storage-trunc}, and individual entries of \(a_h\) can even be negative. Transcript distance requires a bound on a sum of \emph{absolute} probability differences. The following smoothing estimate explains how one more refresh makes that possible. We use \(\|H\|_1=\E_\nu|H|\) and \(\|H\|_{S_4}=(\sum_k\sigma_k(H)^4)^{1/4}\), with the SVD and norm conventions of Appendices~\ref{app:svd} and~\ref{app:norms}.

\begin{lemma}[One round of smoothing]\label{lem:smooth}
Let \(H\) be any real table supported on the encodings of one secret. For the uniform matrix refresh, at any incoming history,
\[
                       \sum_j\ellone{\R_jH}\le C\sqrt K\,q^{1/4}\siv H .
\]
\proofnote{The proof is given in \proofref{pr:storage}{proof:smoothing}. Its short-sampler form is \eqref{eq:general-smoothing}.}
\end{lemma}

There are two ways to get an approximation at \(hj\). We can update \(a_h\) to \(\R_j a_h\), or update the true table and truncate it to \(a_{hj}\). On the encodings of a fixed secret, the cutoff and the history weights control the fourth Schatten norm of their difference. Lemma~\ref{lem:smooth} bounds its total absolute error after the next round. Later rounds can only decrease this error by \eqref{eq:contracts}. The \hyperref[proof:storage]{proof of the storage theorems} adds one contribution \eqref{eq:storage-step} per round and compares each secret to the same law \(h\mapsto m_h\).

The smoothing argument uses a fourth power of a singular-value norm. The four-corner identity \eqref{eq:cycle} expands it into an average over two rows and two columns. This observation is what allows us to shorten the description of \(M\).
\Needspace{9\baselineskip}
\section{Reducing the communication of a refresh}\label{sec:matrix}

Sending the uniform matrix used in Section~\ref{sec:storage} costs \(n^2\) field elements. Suppose instead that the parties sample a matrix from a smaller family and send its description. Which uses of uniformity must that family reproduce? The smoothing estimate \eqref{eq:general-smoothing} needs the images of four indices. The recursion \eqref{eq:history-weights} for the history weights needs the predecessor of one endpoint.

For the first, fix two row indices \(x_1,x_2\) and two column indices \(y_1,y_2\). All four entries after a forward change of basis can be read from the joint tuple
\[
                 Mx_1,\quad Mx_2,\quad M^{-\top}y_1,\quad M^{-\top}y_2 .
\]
We want this tuple to look as it would under a single uniform change of basis, including all four preserved pairings \(x_i^\top y_j\). This is the \emph{forward \((2,2)\) test}. For the history weights, fix just one endpoint \(x,y\) and look at its predecessor \(M^{-1}x,M^\top y\), as in \eqref{eq:round}. Here we want to sample approximately uniformly from the encodings of \(x^\top y\). This is the \emph{reverse \((1,1)\) test}. Forward and reverse refer to the direction in which the proof follows the shares.

\begin{definition}[Images of vectors and covectors]\label{def:mixed}
A sampler has \emph{forward mixed-image error} at most \(\eta\) for \((r,s)\) if, for every fixed list of at most \(r\) vectors \(x_i\) and \(s\) covectors \(y_j\), the joint law of \(((Mx_i)_i,(M^{-\top}y_j)_j)\) has total variation distance at most \(\eta\) from the corresponding law for a uniform matrix in \(\GL_d(\field)\). For the \emph{reverse} error use \(M^{-1}\) in place of \(M\).
\end{definition}

The reference law in Definition~\ref{def:mixed} has the following description in both directions, since the inverse of a uniform matrix in \(\GL_d(\field)\) is also uniform.

\begin{lemma}[Images under a uniform change of basis]\label{lem:orbit}
Fix ordered lists of vectors \((x_i)_i\) and covectors \((y_j)_j\) in dimension \(d\). For uniform \(M\in\GL_d(\field)\), the lists \((Mx_i)_i\) and \((M^{-\top}y_j)_j\) are jointly uniform over all lists \((x'_i)_i,(y'_j)_j\) with exactly the original linear relations within each list and with \((x'_i)^\top y'_j=x_i^\top y_j\) for every \(i,j\). This set of possible images is the \emph{orbit} of the original configuration.
\proofnote{The proof is given in \proofref{pr:sampler}{proof:orbit}.}
\end{lemma}

The comparison in Definition~\ref{def:mixed} concerns the laws of these images. In the protocol both parties see the actual short description, and their leakage functions are always evaluated on that description. The proof only invokes uniform images when estimating the indicated averages.

\topic{A short description}
Split a \(d\)-coordinate vector as \((z_0,z_1)\), with block lengths \(u=\lfloor d/2\rfloor\) and \(v=\lceil d/2\rceil\). A \emph{shear} adds a linear function of one block to the other. For example, \((z_0,z_1)\mapsto(z_0,z_1+A_1z_0)\) is invertible for every rectangular \(A_1\), since we can subtract \(A_1z_0\) to undo it. Using four independent maps, perform the numbered updates in order.
\[
\begin{array}{c@{\quad}l@{\qquad}c@{\quad}l}
 1.&z_1\gets z_1+A_1z_0 & 2.&z_0\gets z_0+A_2z_1\\[2pt]
 3.&z_1\gets z_1+A_3z_0 & 4.&z_0\gets z_0+A_4z_1 .
\end{array}
\]
Each step uses the most recently updated blocks. The composition is one change of basis in one leakage phase. The right party applies its inverse transpose.

The description becomes short by choosing each \(A_i\) from a family of linear maps specified by a few coefficients. Fix a public basis of the extension field \(E=\mathbb F_{q^v}\) over \(\field\). Take \(t\) independent uniform coefficients in \(E\) and form the \(\field\)-linear map
\[
                          A(z)=\sum_{j=0}^{t-1}a_j z^{q^j}.
\]
For a rectangular map \(A_{j,i}\) from \(i\) to \(j\) coordinates, pad the input with zeros to length \(v\), apply \(A\), and keep the first \(j\) coordinates in the chosen basis. Write \(M_t\) for the four updates above when the first map has \(t\) coefficients and the other three have four each. Thus \(M_2\) costs \(14v\) field elements and \(M_1\) costs \(13v\). The former will serve for storage. The latter suffices for the long matrix at a gate, where the proof uses only the forward test.

\begin{lemma}[The four updates]\label{lem:shear}
For \(u\ge4\), both \(M_1\) and \(M_2\) have forward error at most \(Cq^{3-u}\) for \((2,2)\). The reverse error of \(M_2\) for \((1,1)\) obeys the same bound. With four coefficients in every map the bound holds in both directions for \((2,2)\).
\proofnote{The proof is given in \proofref{pr:sampler}{proof:shear}.}
\end{lemma}

Each shear leaves one vector block and the complementary covector block fixed. When the lists in those blocks have full rank, a shear with four coefficients samples the other blocks uniformly subject to the pairing equations. The first update only has to make full rank likely. For either test, suppose one sample has error at most \(\eta\). The error \(\eta_a\) of the product of \(a\) independent samples in a single phase satisfies
\begin{equation}
                       \eta_a\le 2^{a-1}\eta^a\qquad(a\ge1).       \label{eq:repetition}
\end{equation}
\proofnote{The proof is given in \proofref{pr:sampler}{proof:repetition}.}

\Needspace{8\baselineskip}
\begin{theorem}[Storage with a short description]\label{thm:short-storage}
In dimension \(n\ge8\), use an independent \(M_2\) in each storage round. For all \(s,t\in\field\), with \(K=2^{b_L+b_R}\) as in Theorem~\ref{thm:storage},
\begin{equation}
       \TV(P_s^T,P_t^T)\le CT\bigl(K^2q^{\,2-n/8}
                                  +K^{3/2}q^{\,5/2-n/8}\bigr).
       \label{eq:storage-bound}
\end{equation}
The protocol communicates \(14\lceil n/2\rceil\) field elements per round and tolerates every fixed per-party leakage fraction below \(1/32\) with error \(2^{-\Omega(\bitlen)}\) for \(T\) polynomial in \(\bitlen\), as \(n\to\infty\) uniformly in \(q\).
\proofnote{The two storage theorems are proved together in \proofref{pr:storage}{proof:storage}.}
\end{theorem}

The first term in \eqref{eq:storage-bound} is the uniform-matrix bound. \proofanchor{claim:storage-repeat}Two independent short samples per round recover this bound at a cost of \(28\lceil n/2\rceil\) field elements by \eqref{eq:repetition} and the \hyperref[proof:storage-repeat]{two-sample calculation in Section~\ref*{pr:storage}}.
\section{Multiplication and other arithmetic gates}\label{sec:gate}

Suppose the two input wires encode \(s_1=L_1^\top R_1\) and \(s_2=L_2^\top R_2\). Each party can form a tensor using just its own shares. Their inner product already gives the answer
\[
         (L_1\otimes L_2)^\top(R_1\otimes R_2)
                    =(L_1^\top R_1)(L_2^\top R_2)=s_1s_2 .
\]
This is the multiplication step of \cite{DF12}. The problem is to bring the two long vectors back to length \(n\), so the product can be used on an ordinary wire. During that conversion each party may leak from its view. The construction takes two phases, with an erasure between them.

\topic{Phase one changes basis on the long shares}
Write \(x=(L_1,L_2)\), \(y=(R_1,R_2)\), and set \(a_x=L_1\otimes L_2,\ b_y=R_1\otimes R_2\), of length \(m=n^2\). The left party samples and privately sends the description of \(M_1\) in dimension \(m\) (use a uniform dense matrix for \(n<9\)). The parties form
\[
                           U=Ma_x,\qquad V=M^{-\top}b_y .
\]
The left party keeps \(x,U\) and the right party keeps \(y,V\). They erase all other phase-one data, including the matrix description on both sides and the sampling tape on the left. In particular the next phase may leak from the operands and the long shares, using queries chosen after seeing the first answers.

\topic{\proofanchor{claim:shortening}Phase two shortens the shares}
Write \(U=(U_0,U_1)\) and \(V=(V_0,V_1)\). The first blocks have \(n-1\) coordinates, and the tails have \(k=m-n+1\). The goal is to replace the two tails by a single coordinate on each side.
\begin{enumerate}
\item The left party chooses a uniform nonzero scalar \(\alpha\) and privately sends \(Z=U_1/\alpha\). The parties form
\[
                    l=(U_0,\alpha),\qquad r=(V_0,Z^\top V_1).
\]
Their new last coordinates contribute \(\alpha Z^\top V_1=U_1^\top V_1\), exactly the missing part of the long inner product.
\item The right party always returns the private bit \(\ind_{\{r=0\}}\). If it is one, the product is zero and the parties replace \(l,r\) by the distinct basis vectors \(e_1,e_2\). Thus both shares entering the next step are nonzero for every secret.
\item The left party samples and sends a fresh uniform \(Q\in\GL_n(\field)\). The output is \(A=Ql,\ B=Q^{-\top}r\). Only these output shares survive in the gate workspace after the phase.
\end{enumerate}
The messages in these steps, including the flag, belong to the local phase-two views. For every fixed pair entering the last step, \((A,B)\) has exactly law \(\mu_{s_1s_2}\) before the leakage is observed, as shown in the \hyperref[proof:freshness]{freshness argument in Section~\ref*{pr:gate}}. The shortening message consists of one scaled tail. The final uniform matrix acts in the ordinary share dimension \(n\).

\Needspace{8\baselineskip}
\topic{\proofanchor{claim:gate-costs}Other gates and communication}
The same two phases work for the following local preparations. They allow direct arithmetic over the field, and \(1-s_1s_2\) is NAND over \(\mathbb F_2\).
\begin{center}\small
\begin{tabular}{@{}llll@{}}
\toprule
Gate value & Initial left vector \(a_x\) & Initial right vector \(b_y\) & Length \(m\)\\
\midrule
\(s_1s_2\) & \(L_1\otimes L_2\) & \(R_1\otimes R_2\) & \(n^2\)\\
\(c-s_1s_2\) & \((L_1\otimes L_2,1)\) & \((-(R_1\otimes R_2),c)\) & \(n^2+1\)\\
\(s_1+s_2\) & \((L_1,L_2)\) & \((R_1,R_2)\) & \(2n\)\\
\bottomrule
\end{tabular}
\end{center}
The scalar \(c\) is public. When \(c=0\) a change of sign alone suffices, giving \(m=n^2\). For either product gate and \(n\ge9\) the communication is \(13\lceil m/2\rceil+(m-n+1)+n^2\) field elements and one bit. For basic addition use a uniform long matrix, giving \(5n^2+n+1\) field elements and one bit. For \(n\ge19\) one may use five independent short descriptions in that phase, giving \(n^2+66n+1\) field elements and one bit with the same security bound. Section~\ref{pr:gate} gives the \hyperref[proof:implementation]{communication and field-operation counts} and the \hyperref[proof:dimension]{rank argument} showing that exact local preparation for multiplication followed by an inner product needs at least \(n^2\) coordinates.

\Needspace{4\baselineskip}
\topic{What a gate must guarantee to its neighbors}
Freshness explains how to generate the next wire. It cannot, by itself, explain what an answer at a later gate reveals together with an earlier answer. For instance, if \(W\) is a uniform bit, the answers \(W\) and \(W\oplus s\) are each uniform, and together determine \(s\). In a circuit, the random information shared by successive operations comes from their common wire. To compose gates we will keep that wire in the analysis.

Fix the incoming history, the adversary's tape and a complete answer \(z\) of one operation. Write \(\ell_z\) for the probability of this answer \emph{conditional on all the operation's input and output encodings}. At a gate its arguments are the two operand encodings and the short output encoding. We also need a one-phase \emph{input-port refresh}, which applies a uniform matrix in dimension \(n\) to a single encoding. Its arguments are the old and new encoding. Section~\ref{sec:circuit} explains where these refreshes are placed. The earlier table \(p_h\) in \eqref{eq:history} followed the whole past from a current storage endpoint. The table \(\ell_z\) describes just one operation, with all its endpoints specified.

\topic{The first phase of the analysis}
There is a common way to study the long change of basis at a gate and the ordinary change of basis at a port. The old arguments \(x,y\) determine nonzero vectors \(a_x,b_y\). Let \(u,v\) stand for possible new vectors. Consistency requires \(u^\top v=a_x^\top b_y\), with indicator \(J(x,u;y,v)\). For a specified answer in this matrix phase let \(p(x,u;y,v)\) be its probability given those endpoints, extended by zero on unreachable tuples. For now, norms average over independent uniform old local arguments and independent uniform new vectors. Write \(\odot\) for entrywise multiplication.

Why should a long tensor help? There are \(P=(q^n-1)/(q-1)\) one-dimensional subspaces, or \emph{lines}, in \(\field^n\). A uniform share falls on a specified line with probability \(1/P\). A nonzero tensor determines the line of each of its factors. Consequently a uniform pair of operands on one side gives a tensor on a specified line with probability at most \(1/P^2\). Write \(D=P\) for a port and \(D=P^2\) for a gate. This is the source of the extra power of \(P\) in the gate estimate.

\begin{lemma}[Approximation after a matrix phase]\label{lem:phase-app}
For the port and long gate phases described here, retain the singular values of \(p\) exceeding \(D^{-1/3}\), and call the resulting table \(B\). The table \(G=qB\) satisfies
\[
                  \nuc G\le C,\qquad \hs{p-J\odot G}\le CD^{-1/9}.
\]
Here \(\|G\|_*\) is the sum of its singular values and \(\|\cdot\|_2\) is root mean square (Appendix~\ref{app:norms}). The rows are indexed by \((x,u)\) and the columns by \((y,v)\).
\proofnote{The proof is given in \proofref{pr:gate}{proof:phase-app}, using the general estimate proved in \proofref{pr:local}{proof:general-phase}.}
\end{lemma}

The factor \(q\) accounts for the density of the consistency condition, which is about \(1/q\). The balanced SVD factors of Appendix~\ref{app:factors} give a factorization of \(G\) with bounded mean energy on each side. The four-corner estimate used in the proof is \eqref{eq:matrix4}.

\proofanchor{claim:reverse}To transfer this approximation through shortening, think of phase two in reverse. Given a short output \(A,B\) and the transmitted \(Q,Z\), we can recover \(U_0,\alpha\) on the left and \(V_0,Z^\top V_1\) on the right whenever the flag is zero. This determines \(U_1=\alpha Z\). For a uniform long encoding, on the regular event \(Z\ne0,\ r\ne0\), the only vector left to draw is \(V_1\), uniformly on one affine hyperplane. For fixed \(Q,Z\) its defining equation uses just the right-hand data. This is why a product of separate left and right factors can be carried through shortening. Section~\ref{pr:gate} derives this \hyperref[proof:reverse]{conditional law} and uses it to prove \eqref{eq:retained}.

\begin{lemma}[The answer given the wire encodings]\label{lem:retained}
For each history and answer \(z\), there is one table \(G_z=\sum_j r_jt_j\) that works for all valid plaintext labels. Each \(r_j\) uses only the left shares in the operation, and each \(t_j\) only the right shares. On each side, over its independent uniform shares, \(\E\sum_j r_j^2\le C\) and \(\E\sum_j t_j^2\le C\), respectively. For every valid labeling \(\boldsymbol s\),
\begin{equation}
 \|\ell_z-G_z\|_{2,\boldsymbol s}\le
 \begin{cases}
  \delta_R=CqP^{-1/9},        &\text{at an input-port refresh},\\
  \delta_G=Cq^{3/2}P^{-2/9},  &\text{at an arithmetic gate}.
 \end{cases}                                                     \label{eq:retained}
\end{equation}
A labeling is valid if both ends of a refresh carry the same value, or if the gate output is the prescribed function of its inputs. The norm averages the squared error over \emph{independently sampled encodings} of those values and takes the square root.
\proofnote{The proof is given in \proofref{pr:gate}{proof:retained}.}
\end{lemma}

The circuit proof uses this estimate with the wire encodings retained. For a gate used in isolation, integrating them out gives a simpler statement.

\begin{corollary}[Visible leakage of one gate]\label{thm:gate}
For independently encoded operands and any two pairs of plaintext inputs, the distance between the complete leakage transcripts, with the output encoding hidden, is at most \(CK_Gq^{3/2}P^{-2/9}\), where \(K_G\) bounds the number of possible full gate answers. A per-party gate budget \(\lambda\) gives \(K_G\le2^{2\lambda}\), and a per-phase budget \(b\) gives \(K_G\le2^{4b}\). Thus the admissible fractions for exponentially small error are any fixed values below \(1/9\) and \(1/18\), respectively.
\proofnote{The proof is given in \proofref{pr:gate}{proof:gate}.}
\end{corollary}

\section{From gates to stateful circuits}\label{sec:circuit}

A source circuit computes deterministically on private field inputs or private state, public inputs and constants. A trusted encoder samples independent encodings for its initial private wires. A public value is encoded by an ordinary leaky operation in which the left party draws an encoding and sends the right share. A trusted decoder releases the designated plaintext outputs. Between invocations the parties keep the encoded next state. The adversary may use all earlier leakage and authorized outputs to choose new public inputs. The trust assumption concerns input encoding and output decoding. All internal operations, including public samplers and communication of matrix descriptions, obey Section~\ref{sec:model}.

\topic{\proofanchor{claim:wire-schedule}How to read a wire twice}
Suppose a value will be used by two later gates. Call its first encoding \(w_0\). Just before the first use, refresh \(w_0\) to \(w_1\) with a uniform matrix in dimension \(n\), and give the first gate its input copy of \(w_1\). Before the next use, refresh the current encoding \(w_1\) to \(w_2\), and give that copy to the second gate. The first gate and the second refresh are precisely the two operations that read \(w_1\). A wire of any fan-out gives a chain of this kind. Even when both input ports of the \emph{same} gate read one source wire, we use two consecutive refreshes and keep the first port copy inactive during the second.

This schedule has two consequences. Every produced encoding has its required distribution for each fixed input, by Lemma~\ref{lem:orbit} for a refresh and \hyperref[proof:freshness]{output freshness} for a gate. Also, any one encoding occurs in at most one gate and at most two one-phase operations. These bounds hold even for wires retained as state across calls. Counting the two input ports and the gate itself gives \(S=3g+I_{\rm pub}\) internal operations for \(g\) binary gates and \(I_{\rm pub}\) samplers of public inputs or constants.

\Needspace{7\baselineskip}
\begin{theorem}[Arithmetic circuit compilation]\label{thm:circuit}
Over every finite field, the compiled circuit computes the prescribed outputs. For every admissible leakage strategy there is a simulator that receives the public inputs and authorized outputs. For every private initial state the statistical distance between the simulated and real views is \(2^{-\Omega_\rho(\bitlen)}\) as \(n\to\infty\), uniformly in \(q\), for \(S\) polynomial in \(\bitlen\) and either of the following allowances.
\begin{itemize}
\item Each party may leak up to \(\rho\bitlen\) bits per operation, for any fixed \(\rho<1/230\). At a gate the two phase budgets sum to at most this allowance.
\item Each party may leak up to \(\rho\bitlen\) bits in each phase, for any fixed \(\rho<1/305\). A one-phase operation gets this same budget.
\end{itemize}
The simulator is independent of the private state and the target error. With standard field representations, its expected work is polynomial in the size of the unrolled circuit, the total lengths of its inputs and outputs, \(n\), and \(\log q\), in addition to the cost of running the adversary and its leakage functions. The finite bound for arbitrary budgets and operation counts is given in \eqref{eq:circuit-main}.
\proofnote{The proof is given in \proofref{pr:circuit}{proof:circuit}.}
\end{theorem}

\topic{How the local statements fit together}
Fix a private initial state, the adversary's tape and a formal transcript \(z=(z_1,\ldots,z_S)\). Along this transcript each location \(w\) has a prescribed plaintext \(s_z(w)\). In particular, adaptive public inputs are fixed once the earlier answers and authorized outputs are fixed. Let \(X_w\) denote the \emph{pair} of shares at location \(w\) and \(D_i\) all locations used or produced by operation \(i\). Take the actual conditional answer tables \(\ell_{i,z_{<i},z_i}\) from Section~\ref{sec:gate}, including the exact conditional probability for a public sampler. Then
\begin{equation}
 \Prb[Z=z]=
   \mathop{\E}\limits_{\substack{X_w\sim\mu_{s_z(w)}\\\text{independently for all }w}}
       \left[\prod_{i=1}^{S}\ell_{i,z_{<i},z_i}(X_{D_i})\right].
       \label{eq:canonical}
\end{equation}
For example, at \(w_1\) in the two-use chain we draw one encoding and use that \emph{same} draw in each incident factor. The \hyperref[proof:canonical]{derivation of (\ref*{eq:canonical})} factors each operation's joint law into its fresh output law and its answer probability given the endpoints. The independent draws in \eqref{eq:canonical} compute the probability of a specified transcript. After observing that transcript, the wire encodings can of course be correlated.

For a common reference expression, use the approximations from Lemma~\ref{lem:retained} at gates and ports, and two independent uniform shares at every location except a public sampler output. Keep the public samplers exact. The expansion \eqref{eq:marked-expansion} recovers \eqref{eq:canonical} as its unmarked term. Wherever a wire's law changes, it keeps the separate left and right factors on every operation touching that wire. A local gate or port error can then occur only on wires with their true encoding laws. The input-port schedule bounds the number of neighbors in either case. The product-space H\"older inequality \eqref{eq:finner} controls the products of errors on shared wires. Counting neighbors gives the three costs in \eqref{eq:marked-cost}. Balancing them in \eqref{eq:exponents} gives the two leakage rates of Theorem~\ref{thm:circuit}.

\proofanchor{claim:simulator}The simulator itself has a short description. Run the compiled circuit on a dummy private state to answer the leakage queries. At each invocation give the adversary the true authorized output received from the ideal computation. For a fixed answer history, the real and dummy executions make the same public choices. The \hyperref[proof:simulation]{simulation argument in Section~\ref*{pr:circuit}} applies the distance bound \eqref{eq:comp-general} to the entire interaction, including all uses of the persistent state.
\Needspace{9\baselineskip}
\section{Technical Proofs}\label{sec:proofs}

In this section we prove the storage and circuit theorems. We first establish the properties of the short matrix sampler and use them to compare the leakage transcripts for two stored secrets (Sections~\ref{pr:sampler} and~\ref{pr:storage}). For circuits, we approximate the probability of a leakage answer at one gate given its input and output shares (Sections~\ref{pr:local} and~\ref{pr:gate}). Section~\ref{pr:circuit} combines these approximations across shared wires and derives the compiler's leakage rates.

All vector averages in a table norm are with respect to the indicated uniform product measures, unless a fixed labeling is displayed. Constants denoted by \(C\) are absolute and may change between inequalities.

\subsection{Orbits and the four-shear sampler}\label{pr:sampler}

\begin{proof}[\proofanchor{proof:orbit}Proof of Lemma~\ref{lem:orbit} (uniform changes of basis)]
Every change of basis preserves the stated relations and pairings. Conversely, fix target lists satisfying these conditions. The rule \(x_i\mapsto x'_i\) defines an isomorphism \(M_0:X\to X'\) between the spans of the two vector lists. Select a basis from the covector list. The same positions give a basis in the target list. Let \(B,B':\field^d\to\field^k\) evaluate the respective \(k\) selected covectors. The pairings give \(B'M_0=B|_X\), so \(M_0\) maps \(X\cap\ker B\) onto \(X'\cap\ker B'\).

Both evaluation maps are onto. Their kernels have equal dimension, so extend the map on the intersection to an isomorphism \(\ker B\to\ker B'\) by completing bases. Together with \(M_0\) this defines \(M\) on \(X+\ker B\). For each basis vector of a complement of \(B(X)=B'(X')\) in \(\field^k\), choose one preimage under each evaluation map and let \(M\) send the first to the second. Now \(M\) is invertible on the whole space and \(B'M=B\). Thus \(M^{-\top}y_j=y'_j\) for the selected covectors and, by their linear relations, for all \(j\).

Every prescribed image is therefore reachable. Composition with a change of basis carrying one image to another gives a bijection between their sets of preimages in \(\GL_d(\field)\). A uniform matrix therefore gives them equal probability.
\end{proof}

\begin{proof}[\proofanchor{proof:shear}Proof of Lemma~\ref{lem:shear}]
We first record exactly what a random rectangular map supplies. For \(t\) coefficients and \(a\le t\) independent arguments in \(E\), their images under \(A\) are independent and uniform in \(E\). Indeed, the evaluation matrix of the first \(a\) Frobenius powers is nonsingular. Otherwise a nonzero linearized polynomial of ordinary degree at most \(q^{a-1}\) would vanish on the span of the arguments, which has \(q^a\) elements. This is the usual Moore-matrix argument \cite{LN,Gab85}. For a rectangular restriction \(B=A_{j,i}\), the combined observations \(BX,B^\top Y\), for lists of lengths \(r,s\) with \(r+s\le t\), have the same law as for a uniform \(j\)-by-\(i\) matrix. A linear test of all observed entries is a trace pairing with a coefficient matrix of rank at most \(r+s\). Extending this matrix by zeros keeps its rank. A rank factorization expresses the test as a sum of linear functionals of \(A\) on at most \(t\) independent arguments. Thus the same linear tests vanish identically in both experiments. The two observation spaces are equal, since a subspace is determined by the linear functionals vanishing on it. Each experiment is uniform on its image.

Let \(r,s\le2\) now denote the ranks of the two lists. It suffices to track bases. Split their column matrices into top and bottom blocks \(X_0,X_1\) and \(Y_0,Y_1\). Keep the two half-configurations
\[
    H=(X_1,Y_0),\qquad J=(X_0,Y_1),\qquad
    J_X^\top H_Y+H_X^\top J_Y=C_0 ,
\]
where \(C_0\) is the invariant pairing matrix. Call a half full if its two matrices have column ranks \(r,s\). The numbers of unrestricted half-configurations are
\(B_H=q^{vr+us}\) and \(B_J=q^{ur+vs}\). A lower shear changes only \(H\), and an upper shear changes only \(J\).

The first lower step leaves \(H\) full except with probability \(\epsilon\le(q^r+q^s)q^{-u}\). For a nonzero coefficient vector \(a\), either \(X_0a=0\) and \(X_1a\ne0\), or \(X_1a+B X_0a\) is uniform. The same test on \(Y_0-B^\top Y_1\) gives the second term. Only a single ordinary or transposed image of \(B\) is used at a time, so one coefficient suffices at this step.

From a fixed full \(H\), an upper step draws \(J\) uniformly from its \(F_J=B_Jq^{-rs}\) compatible values. In suitable bases prescribing \(BH_X\) and \(-B^\top H_Y\) fixes \(r\) columns and \(s\) rows of a rectangular matrix, with \(rs\) entries shared. Thus every compatible prescription has equally many preimages, and the preceding observation permits the short description. From a full \(J\) the lower step similarly draws \(H\) uniformly from \(F_H=B_Hq^{-rs}\) values.

Fix a full \(H\) after step one, and draw a candidate \(H'\) uniformly from the unrestricted half space. Except with probability \(Cq^{3-u}\), both concatenations \([H_X,H'_X]\), \([H_Y,H'_Y]\) have full column rank. In adding \(r\le2\) columns to an existing \(r\)-dimensional span in dimension \(v\), the failure probability is at most \(\sum_{a=r}^{2r-1}q^{a-v}\). The other concatenation has the same bound with \(s,u\). For such \(H'\) there are \(B_Jq^{-2rs}\) values of \(J\) satisfying the two pairing equations for \(H\) and \(H'\). Each of \(J_X,J_Y\) has a uniform marginal on this intersection. For example, once \(J_X\) is fixed the two equations on \(J_Y\) have rank \(2rs\) by the concatenation condition. Thus at least a \(1-\epsilon\) fraction of these \(J\)'s are full. Following just the paths through full \(J\)'s gives
\[
 \Prb[H_{\rm after\ step\ 3}=H'\mid H_{\rm after\ step\ 1}=H]
 \ \ge\ (1-\epsilon)\frac{B_Jq^{-2rs}}{F_J F_H}
 =\frac{1-\epsilon}{B_H}.
\]
Summing the overlap with the uniform distribution shows that the step-three half is \(Cq^{3-u}\)-close to uniform, also after averaging over the outcome of step one.

By Lemma~\ref{lem:orbit}, the orbit now consists of all configurations in which each list is linearly independent and the pairing matrix is \(C_0\). Every full \(H\) ensures this independence and has exactly \(F_J\) compatible completions. Thus the orbit law gives equal mass to every full \(H\). The probability of a nonfull half on that orbit is at most \(\epsilon\), since every nonzero combination within either original list has a uniform nonzero image. The same bound applies to an unrestricted uniform half. Step four draws precisely the orbit conditional law of \(J\) on every full \(H\). This proves the forward bound. In reverse order upper and lower interchange and the signs change. The last reverse step is the first forward map. For a single vector and covector it requires only two coefficients, and for two of each it requires four. A coefficient costs \(v\) base-field symbols. This proves all assertions.
\end{proof}

\begin{proof}[\proofanchor{proof:repetition}Proof of \eqref{eq:repetition} (independent samples)]
Consider the transition kernel induced on any orbit of the test in Definition~\ref{def:mixed}. Uniform measure is stationary since every matrix permutes that orbit. If each row has distance at most \(\eta\) from uniform, two rows have distance at most \(2\eta\). Applying the kernel to two measures contracts their distance by a factor at most \(2\eta\), by coupling or by separating their common mass. Iterating from a fixed point gives \(2^{a-1}\eta^a\). The same reasoning applies to a reverse test when its hypothesis holds.
\end{proof}

\subsection{The storage bound}\label{pr:storage}

\begin{proof}[\proofanchor{proof:smoothing}Proof of Lemma~\ref{lem:smooth} and its extension \eqref{eq:general-smoothing}]
Fix an incoming history and a real table \(H\) supported on \(\orb n s\). Put \(K=2^{b_L+b_R}\) as in Theorem~\ref{thm:storage}. If the sampler has forward \((2,2)\) error at most \(\eta\) (Definition~\ref{def:mixed}), the round operators in \eqref{eq:round} satisfy
\begin{equation}
  \sum_j\ellone{\R_jH}\le C\sqrt K
       \bigl(q^{1/4}\siv H+\eta^{1/4}\hs H\bigr).
  \label{eq:general-smoothing}
\end{equation}
For a uniform matrix \(\eta=0\), giving precisely Lemma~\ref{lem:smooth}. The second term accounts for the short sampler in Theorem~\ref{thm:short-storage}.

For a bounded test \(|d|\le1\) let \(d_M(x,y)=d(Mx,M^{-\top}y)\). First take \(M\) uniform on \(\GL_n(\field)\). In the four-corner formula \eqref{eq:cycle} for \(\siv{H\odot d_M}^4\) let \(A\) be the product of the four entries of \(H\), and \(B\) the average over \(M\) of the product of the four entries of \(d_M\). Whenever \(A\ne0\), all four pairings are \(s\). If each list is linearly independent, Lemma~\ref{lem:orbit} gives the same image law for every corner with these four pairings. Let \(b_0\in[-1,1]\) be this common value of \(B\), taking \(b_0=0\) if there are no such corners.

It remains to bound exceptional corners without losing cancellation in \(\E A=\siv H^4\). For fixed \(\lambda\in\field^\times\) write \(r(x)=\E_y H(x,y)^2\). Then
\[
 \E[\ind_{\{x'=\lambda x\}}|A|]
 \le N^{-1}\E_x r(x)r(\lambda x)
 \le N^{-1}\E_x r(x)^2\le\siv H^4 .
\]
The first inequality averages over the two columns, the second uses a permutation, and the third retains the diagonal in the nonnegative sum of squared row correlations in \eqref{eq:cycle}. The same bound holds on each exceptional column event. In \(\E[AB]=b_0\E A+\E[A(B-b_0)]\), the latter costs at most \(2\) on each of \(2(q-1)\) exceptional events. Consequently the uniform average of \(\siv{H\odot d_M}^4\) is at most \(4q\siv H^4\). For the actual sampler the average of any fixed product of four test entries changes by at most \(2\eta\) by \eqref{eq:tv-test}. Its total additional cost is at most \(2\eta\,\E|A|=2\eta\,\siv{|H|}^4\le2\eta\,\hs H^4\) by \eqref{eq:schatten}.

Take \(d_j=\operatorname{sgn}(\R_jH)\). Changing variables in \eqref{eq:round} expresses \(\sum_j\ellone{\R_jH}\) as the expectation over \(\xi\) of the sum of tests of \(H\odot(d_j)_M\) against the two local indicators for leaf \(j\). For fixed \(\xi\) the sum of the leaf masses is one by the \hyperref[claim:rectangles]{rectangle partition in Section~\ref*{sec:storage}}. The rank-one test bound from \eqref{eq:duality} and Cauchy--Schwarz bound the sum of tests by \((\sum_j\op{H\odot(d_j)_M}^2)^{1/2}\). Average, apply Jensen, and bound each mean squared operator norm by the square root of the fourth-moment bound just obtained. There are at most \(K\) terms, giving \eqref{eq:general-smoothing}.
\end{proof}

\begin{proof}[\proofanchor{proof:storage}Proof of Theorems~\ref{thm:storage} and~\ref{thm:short-storage}]
We first estimate the error of truncating one history table on a fixed secret. We then apply the smoothing bound \eqref{eq:general-smoothing} to add these errors over successive rounds.

\partitle{The encoding density and the SVD cutoff}
Write \(J_{d,s}(x,y)=\ind_{\{x^\top y=s\}}\) and let \(\pi_{d,s}\) be its mean under \(\nu_d\). For each nonzero row the pairing equation has \(q^{d-1}\) solutions before excluding the zero vector, and the same count holds for every column. The \hyperref[claim:encoding]{encoding distribution of Section~\ref*{sec:model}} therefore satisfies
\begin{equation}
 \pi_{d,s}=\frac{q^{d-1}-\ind_{\{s=0\}}}{q^d-1}\ge\frac{1}{2q}
 \quad(d\ge2),\qquad
 \frac{d\law d s}{d\nu_d}=\frac{J_{d,s}}{\pi_{d,s}}.             \label{eq:orbit-density}
\end{equation}
Both the storage and gate proofs use the hyperplane bound in Appendix~\ref{app:incidence}. Write \(\Phi_{d,s}=J_{d,s}-\pi_{d,s}\). With uniform row and column indices on \(\Omega_d\), \eqref{eq:hyperplane} bounds the norm of \(J_{d,s}-1/q\) by \((q-1)q^{d/2}/(qN_d)\). The rows and columns of \(J_{d,s}\) all have mean \(\pi_{d,s}\), so projecting \(J_{d,s}-1/q\) onto the mean-zero functions on both sides gives \(\Phi_{d,s}\). Therefore
\begin{equation}
            \op{\Phi_{d,s}}\le q^{d/2}/N_d\le 2q^{-d/2}.
            \label{eq:orbit-mask}
\end{equation}

Apply the \hyperref[claim:history-trunc]{cutoff of Section~\ref*{sec:storage}} to a table \(p\) with \(\supn p=w>0\), writing \(a\) for its truncation at \(\theta w\). We have
\begin{equation}
 \op{p-a}\le\theta w,\qquad \hs a\le\hs p,\qquad
 \op{\Phi_{n,s}\odot a}\le 2\theta w,\quad \theta=q^{-n/4}.
 \label{eq:storage-trunc}
\end{equation}
The first two bounds follow directly from the SVD (Appendix~\ref{app:svd}). Each row and column of \(p\) has mean square at most \(w^2\). With cutoff \(\theta w\), the truncation bounds \eqref{eq:truncation-energies} give pointwise energy at most \(w/\theta\) for the balanced factors of \(a\). The last inequality follows from \eqref{eq:schur} and \eqref{eq:orbit-mask}. The bounds also hold when \(w=0\) and \(a=0\).

\partitle{History weights and one-step errors}
Let \(\eta\) bound the forward \((2,2)\) and reverse \((1,1)\) errors of Definition~\ref{def:mixed}. For each endpoint with product \(s\), summing \eqref{eq:round} over all leaves gives the average of \(p_h\) on the reverse image of that endpoint. Its uniform-orbit average is \(P_s^i(h)\) by \eqref{eq:history}. Since \(0\le p_h\le w_h\), \eqref{eq:tv-test} gives
\begin{equation}
       w_{hj}\le\max_sP_s^i(h)+\eta w_h,\qquad
       W_{i+1}\le K(q+\eta W_i),\qquad W_0=1.                         \label{eq:history-weights}
\end{equation}
In particular \(W_i\le2Kq\) if \(K\eta\le1/2\). For the short sampler Lemma~\ref{lem:shear} allows \(\eta=\min(1,C_0q^{3-\lfloor n/2\rfloor})\le C_0q^{4-n/2}\). If \(K\eta>1/2\), \(K\ge (2C_0)^{-1}q^{n/2-4}\) and \eqref{eq:storage-bound} already dominates one for \(T\ge1,n\ge8\) because \(7n/8-6\ge0\). Here the absolute constant may be increased. The uniform sampler has \(\eta=0\).

At each fixed history the round satisfies
\begin{equation}
 \op{\R_jH}\le\op H,\qquad
 \sum_j\hs{\R_jH}^2\le\hs H^2,\qquad
 \sum_j\ellone{\R_jH}\le\ellone H .
 \label{eq:contracts}
\end{equation}
For each fixed sampling tape the map composes with two permutations and restricts to a rectangle. Row and column restrictions are contractions in operator norm. The rectangles partition the product for the last two inequalities, which persist under averaging by convexity.

Fix \(s\), set \(J=J_{n,s},\ \pi=\pi_{n,s}\), and put \(F_h=J\odot a_h-\pi p_h\). The difference between updating a truncation and truncating the updated table on this orbit is
\begin{equation}
 D_{hj}:=J\odot(\R_j a_h-a_{hj})=\R_jF_h-F_{hj},              \label{eq:defect}
\end{equation}
since \(p_{hj}=\R_jp_h\) and \(J\) is invariant. Writing \(F_h=(J-\pi)\odot a_h+\pi(a_h-p_h)\) gives \(\op{F_h}\le3\theta w_h\) by \eqref{eq:storage-trunc}. Also \(\hs{F_h}\le2\hs{p_h}\le2\sqrt{w_hm_h}\). The interpolation bound \eqref{eq:schatten} and the contractions \eqref{eq:contracts} give
\[
 \sum_j\siv{\R_jF_h}\le
 C K^{3/4}\sqrt\theta\,w_h^{3/4}m_h^{1/4},\qquad
 \sum_j\hs{\R_jF_h}\le2\sqrt{K w_hm_h}.
\]
For \(F_h\) itself the same inequalities hold without the powers of \(K\). For example, the first bound uses \(\sum_j(\hs{\R_jF_h}^2)^{1/4}\le K^{3/4}(\sum_j\hs{\R_jF_h}^2)^{1/4}\).

Sum \eqref{eq:defect} over a level of histories and its extensions. H\"older, \(\sum_hm_h=1\) and \(W_i,W_{i+1}\le2Kq\) give
\[
 \sum_{h,j}\siv{D_{hj}}\le
 C\sqrt\theta\,(K^{3/4}+1)(Kq)^{3/4},\qquad
 \sum_{h,j}\hs{D_{hj}}\le C(\sqrt K+1)\sqrt{Kq}.
\]
Each \(D_{hj}\) is supported on the orbit. For every level \(i\), smoothing each such difference once by \eqref{eq:general-smoothing} gives
\begin{equation}
 \sum_{|h|=i}\sum_{j,k}\ellone{\R_{hj,k}D_{hj}}
 \le \varepsilon:=C\bigl(K^2q\sqrt\theta+K^{3/2}q^{1/2}\eta^{1/4}\bigr).
 \label{eq:storage-step}
\end{equation}

\partitle{Accumulation over rounds}
Measure the discarded part of each true table \emph{after one further round}. Put
\[
 E_i=\sum_{|h|=i}\sum_j\ellone{\R_{h,j}[J\odot(p_h-a_h)]}.
\]
We have \(E_0=0\) since \(p_\varnothing=a_\varnothing=1\). Invariance of \(J\) gives
\[
 J\odot(p_{hj}-a_{hj})=\R_{h,j}[J\odot(p_h-a_h)]+D_{hj}.
\]
After the next round, the first term contributes at most \(E_i\) by \eqref{eq:contracts} and the second at most \(\varepsilon\) by \eqref{eq:storage-step}. Thus \(E_{i+1}\le E_i+\varepsilon\), giving \(E_{T-1}\le(T-1)\varepsilon\) for \(T\ge1\).

For the last round, \(|h|=T-1\), compare directly with the probability \(m_{hj}\) in the product experiment. By \eqref{eq:history} and the definition of \(F_h\),
\[
 \pi\bigl(P_s^T(hj)-m_{hj}\bigr)
 =\E_\nu\R_{h,j}[J\odot(p_h-a_h)]+\E_\nu\R_{h,j}F_h .
\]
For a fixed sampling tape the second expectation tests \(F_h\) on the leaf rectangle \(A_j\times B_j\), at cost at most \(\op{F_h}\sqrt{\nu(A_j\times B_j)}\) by \eqref{eq:duality}. The rectangles partition the space, so the sum of these square roots is at most \(\sqrt K\). Averaging over the tape and using \(\op{F_h}\le3\theta w_h\) yields
\[
 \pi\sum_{|z|=T}|P_s^T(z)-m_z|\le (T-1)\varepsilon+3\sqrt K\,\theta W_{T-1}.
\]
The law \(m\) is common to both secrets, so the triangle inequality compares them. Using \(\pi\ge1/(2q)\), \(W_{T-1}\le2Kq\), \(\sqrt\theta=q^{-n/8}\) and the stated sampler errors gives both theorems. For \(T=0\) the transcript is empty.
\end{proof}

\partitle{\proofanchor{proof:storage-repeat}The \hyperref[claim:storage-repeat]{two-sample storage bound in Section~\ref*{sec:matrix}}}
Two independent short descriptions cost \(28\lceil n/2\rceil\) symbols. If one sample has error \(\eta\le Cq^{4-n/2}\) in both required tests, \eqref{eq:repetition} bounds the two-sample error by \(2\eta^2\). Substituting this in \eqref{eq:storage-step} and making the same final comparison gives \(CTK^2q^{2-n/8}\) for \(n>16\). For \(8\le n\le16\) and \(T\ge1\) this bound already dominates one.

\Needspace{12\baselineskip}
\subsection{The answer table of a matrix phase}\label{pr:local}

The following general form of Lemma~\ref{lem:phase-app} covers both a port and a long gate phase. Section~\ref{pr:gate} \hyperref[proof:phase-app]{checks its hypotheses} for the actual samplers.

\begin{lemma}[General matrix phase]\label{lem:local}
Let \(x,y\) be independent uniform old local arguments with nonzero vectors \(a_x,b_y\in\Omega_d\), and let \(u,v\) be independent uniform new vectors in \(\Omega_d\), independent of the old arguments. Set \(J(x,u;y,v)=\ind_{\{u^\top v=a_x^\top b_y\}}\). For some \(D\ge2\), suppose every line has mass at most \(1/D\) under each of \(a_x,b_y\). Suppose the sampler has forward \((2,2)\) error at most \(C_0D^{-20/9}\). For a fixed leaf let \(p(x,u;y,v)\) be its probability given the old arguments \(x,y\) and new vectors \(u,v\), extended by zero on unreachable tuples, and let \(B\) retain its singular values exceeding \(\beta=D^{-1/3}\). Then \(G=qB\) satisfies
\[
                 \nuc G\le C,\qquad \hs{p-J\odot G}\le CD^{-1/9}.
\]
Its balanced factors have pointwise energies at most \(2/\beta\) on each side. Here \(C\) depends only on the absolute constant \(C_0\).
\proofnote{The proof is given in \proofref{pr:local}{proof:general-phase}.}
\end{lemma}

\begin{proof}[\proofanchor{proof:general-phase}Proof of Lemma~\ref{lem:local}]
Let \(\eta\le C_0D^{-20/9}\) be the mixed-image error. The line bound implies \((q-1)/N_d\le 1/D\), since the \(N_d/(q-1)\) lines carry total mass one. For the consistency mask, write \(R=(a_x,u)\) and \(Z=(-b_y,v)\), so \(J=\ind_{\{R^\top Z=0\}}\). Under the reference product measure, every atom of either index has mass at most \(1/(D N_d)\), since the new vector is independent and uniform. In dimension \(2d\), the hyperplane bound \eqref{eq:hyperplane} gives
\begin{equation}
                  \op{qJ-1}\le\frac{(q-1)q^d}{D N_d}\le\frac{2q}{D}.                     \label{eq:phase-mask}
\end{equation}

For a real test \(W\), write \(W^M(x,y)=W(x,Ma_x;y,M^{-\top}b_y)\). Fix the full matrix-sampling tape \(\xi\), as in Section~\ref{sec:model}. For the specified leaf the local answer indicators are \(f_\xi(x)\) and \(g_\xi(y)\). The right-hand indicator depends on \(\xi\) only through the transmitted description \(d(\xi)\). We have
\begin{equation}
 \left|\langle p,W\rangle-q^{-1}\E_{\xi,x,y}f_\xi(x)g_\xi(y)W^M(x,y)\right|
 \le C\bigl(\eta/q+1/N_d\bigr)\supn W.                    \label{eq:dual}
\end{equation}
Indeed, at fixed \(x,y\) the independent \(u,v\) fall on the orbit of \(a_x^\top b_y=s\) with probability \(\pi_{d,s}\). By \eqref{eq:tv-test} the uniform orbit mean of \(pW\) differs from the actual image mean by at most \(2\eta\supn W\). Under the latter distribution the defining conditional expectation of \(p\) recovers the actual answer indicators. Finally \eqref{eq:orbit-density} gives \(|\pi_{d,s}-1/q|\le1/N_d\). This comparison allows the answers to depend on any sampling coins in the respective local views.

\partitle{Mean energy of the truncation}
We first prove the nuclear-norm bound in Lemma~\ref{lem:local}. Write \(B=\sum_{\sigma_j>\beta}\sigma_j e_jh_j^\top\) and let \(r\) be the number of kept terms. If \(r=0\), its energy is zero and the residual argument below still applies. Since \(0\le p\le J\), the squared \(L^2\) norm of each row and column is at most \(2/q\). For example, for fixed \(x,u,y\), the proportion of compatible \(v\)'s is \(\pi_{d,a_x^\top b_y}\). The truncation bounds \eqref{eq:truncation-energies}, with cutoff \(\beta\) and mean square \(2/q\), give
\begin{equation}
 \enL(G),\,\enR(G)\le 2/\beta,\qquad
 \|\boldsymbol e\|^2,\,\|\boldsymbol h\|^2\le L_0:=2/(q\beta^2),                                          \label{eq:phase-energies}
\end{equation}
where \(\boldsymbol e,\boldsymbol h\) collect the kept singular functions. Here \(\enL,\enR\) denote the maximum row and column energies for the factors \(\sqrt{q\sigma_j}e_j,\sqrt{q\sigma_j}h_j\).
Set \(A=q\nuc B\), so \(r\le A/(q\beta)\).

For a uniform matrix \(M\), and Hilbert-space-valued functions \(k_x(u)\), one has the variance bound
\begin{equation}
 \E_M\left\|\E_x k_x(Ma_x)-\E_{x,u}k_x(u)\right\|^2
 \le (C/D)\E_{x,u}\|k_x(u)\|^2.                            \label{eq:gram}
\end{equation}
Center each function in \(u\) and expand the square. For proportional \(a_x,a_{x'}\) the absolute covariance is at most the product of their centered \(L^2\) norms. The contribution averaged over these pairs is at most \(D^{-1}\) times the mean squared norm, by \(2ab\le a^2+b^2\) and the mass bound for each line. Otherwise the pair of images is uniform on nonproportional vectors by Lemma~\ref{lem:orbit}. Two independent uniform nonzero vectors are proportional with probability \(\rho_0=(q-1)/N_d\). On that event each marginal remains uniform. Since the unconditional centered covariance is zero, conditioning on its complement costs at most \(\rho_0/(1-\rho_0)\) times the product of the centered norms. As \(\rho_0\le 1/D\le1/2\), this proves \eqref{eq:gram}. The dual vectors obey the same bound.

Apply \eqref{eq:gram} to \(k_x(u)=\boldsymbol e(x,u)\boldsymbol e(x,u)^\top\) with Frobenius norm. Its mean is \(I_r\), and \(\E\|k_x(u)\|_{\rm F}^2\le L_0 r\). Let \(E_\xi=\E_x k_x(Ma_x)\). In expanding \(\E\|E_\xi-I_r\|_{\rm F}^2\) we use one- and two-image averages. Replacing a uniform matrix with the actual sampler costs \(C\eta(L_0^2+L_0)\) by \eqref{eq:tv-test}, since a pair term is bounded by \(L_0^2\) and the linear term tests \(\tr k_x(u)\le L_0\). For \(r>0\), \(L_0\ge1\). Thus
\[
 \E\op{E_\xi}\le
  1+C\sqrt{L_0r/D}+C\sqrt{\eta}\,L_0 .
\]
The right Gram matrix satisfies the same inequality. In \eqref{eq:dual} use \(W=\sum_{j=1}^r e_jh_j^\top\), for which \(\langle p,W\rangle=\nuc B\) and \(\supn W\le L_0\). For fixed \(\xi\), the vector \(\E_x f_\xi(x)\boldsymbol e(x,Ma_x)\) has squared norm at most \(\op{E_\xi}\) by testing on a unit vector. The same holds on the right, so Cauchy--Schwarz over \(\xi\) gives
\[
 A\le 1+C\sqrt{L_0r/D}+C\sqrt\eta\,L_0+C(\eta+q/N_d)L_0
   \le C+C\sqrt A .
\]
Here \(D\beta^3=1\), \(N_d\ge D\), and \(r\le A/(q\beta)\). This proves \(A\le C\). Keeping the actual rank in the Gram estimate allows the threshold \(D^{-1/3}\).

\partitle{The approximation error}
For the second assertion of Lemma~\ref{lem:local}, put \(H=p-J\odot G,\ t=\hs H,\ B_\infty=\supn H\), and let \(B_2\) be the largest row or column \(L^2\) norm of \(H\). In \(H=(p-B)-(qJ-1)\odot B\), the first term is the SVD remainder. Apply \eqref{eq:schur} to the second term using the mask bound \eqref{eq:phase-mask} and the energies \eqref{eq:phase-energies}. This gives
\begin{equation}
 \op H\le\beta+4/(D\beta)\le C\beta,\qquad B_\infty\le C/\beta,\qquad B_2\le C\sqrt q.
 \label{eq:size-residual}
\end{equation}
For the last bound, truncation can only decrease the mean square of an individual row or column by \eqref{eq:row-pythagoras}. Since \(H\) vanishes off \(J=1\), operator--nuclear duality \eqref{eq:duality} and \(\nuc G=A\le C\) give
\begin{equation}
 t^2=\langle p,H\rangle-\langle G,H\rangle
       \le |\langle p,H\rangle|+\nuc G\,\op H
       \le |\langle p,H\rangle|+C\beta .                     \label{eq:residual-duality}
\end{equation}

\partitle{Four corners with prescribed rows and columns}
For a uniform \(M\), the estimate we need to control \(\langle p,H\rangle\) in \eqref{eq:residual-duality} is
\begin{equation}
 \E_M\cyc(H^M)\le Cq^4\left\{\cyc(H)
       +\left(\frac{B_2^2}{q^2D}+\frac{B_\infty^2}{D^2}\right)t^2\right\},
 \qquad \cyc(H)=\siv H^4.                                  \label{eq:matrix4}
\end{equation}
To estimate the four-corner average, we prescribe row and column images of the same matrix. Transpose the uniform matrix and take an auxiliary \(Q_0\) uniform over \emph{all} \(d\)-by-\(d\) matrices. Restrict \(H\) by multiplying it by \(\ind_{\{a_x^\top Q_0=u^\top\}}\ind_{\{Q_0v=b_y\}}\), obtaining \(H_{Q_0}\). For invertible \(Q_0\), \(N_d^4\cyc(H_{Q_0})=\cyc(H^{Q_0^\top})\) because the old arguments specify the four new vectors. For every \(Q_0\) the full four-corner average is nonnegative by \eqref{eq:cycle}. Since a uniform matrix is invertible with probability \(\prod_{j=1}^d(1-q^{-j})>1/4\), it suffices to bound \(4N_d^4\E\cyc(H_{Q_0})\).

Choose two old left arguments \(x_1,x_2\), two new left vectors \(u_1,u_2\), and independently their right counterparts \(y_1,y_2,v_1,v_2\). Write \(a_i=a_{x_i}, b_j=b_{y_j}\), and \(H_{ij}=H(x_i,u_i;y_j,v_j)\). Let \(r_0=\rank(a_1,a_2)\), \(s_0=\rank(v_1,v_2)\). If prescriptions are consistent, their probability for a uniform \(Q_0\) is \(q^{-(r_0+s_0)d+r_0s_0}\), from specifying \(r_0\) rows and \(s_0\) columns in suitable bases. The four cross-pairing equations hold on the support of \(\prod H_{ij}\). When a list is dependent, its image must also have the same linear relation.
\begin{itemize}
\item If \((r_0,s_0)=(2,2)\) the coefficient after scaling by \(N_d^4\) is at most \(q^4\). Remove from the full signed cycle the event of proportional old rows and the event of proportional new columns by inclusion--exclusion. Each separately subtracted average is nonnegative. At fixed rows the average over the two columns is a squared row correlation, and conversely at fixed columns. On the intersection of the events the conditional probability given any one corner is at most \(D^{-1}(q-1)/N_d\le D^{-2}\). Also \(|\prod H_{ij}|\le B_\infty^2\sum_{i,j}H_{ij}^2/4\). The contribution is therefore at most \(q^4(\cyc(H)+B_\infty^2t^2/D^2)\).
\item For \((1,2)\), the old rows must be proportional and \(u_2\) must be the same multiple of \(u_1\). Given the first full row this costs at most \(1/(D N_d)\). The coefficient is at most \(q^2N_d\). At fixed rows, averaging the absolute product over independent columns is bounded by the product of the two squared row norms. Bound one by \(B_2^2\), and average the first row's squared norm to \(t^2\). The contribution is at most \(q^2 B_2^2 t^2/D\). For \((2,1)\) the analogous bound follows by first requiring proportional old \(b_1,b_2\) and then the same multiple for \(v_1,v_2\).
\item At \((1,1)\) the two consistent proportionalities cost at most \((D N_d)^{-2}\) conditional on any corner. The coefficient is at most \(qN_d^2\), and the bound by the four corner squares gives \(qB_\infty^2t^2/D^2\).
\end{itemize}
These estimates give \eqref{eq:matrix4}. In the first case one can apply the coefficient bound after bounding the signed average by the displayed nonnegative quantity, so negative corner contributions cause no loss of cancellation.

For the actual sampler, the forward test and \eqref{eq:tv-test} change the average cycle by at most \(2\eta B_\infty^4\). By \eqref{eq:schatten}, \(\cyc(H)\le\op H^2t^2\), so \eqref{eq:size-residual} and \eqref{eq:matrix4} imply
\[
        \E_\xi\op{H^M}\le Cq\sqrt{\beta t}+C\eta^{1/4}/\beta.
\]
With \(W=H\), the two local indicators in \eqref{eq:dual} test \(H^M\) at cost at most its operator norm by \eqref{eq:duality}. The sampler term satisfies \(\eta^{1/4}/\beta\le C\beta^{2/3}\), since \(D^{-20/9}=\beta^{20/3}\). Using \eqref{eq:residual-duality}, \(1/N_d\le 1/D=\beta^3\) and \(\beta\le\beta^{2/3}\), we get \(t^2\le C\sqrt{\beta t}+C\beta^{2/3}\). In either case of the larger summand, \(t\le C\beta^{1/3}=CD^{-1/9}\), as asserted.
\end{proof}

\subsection{From the long matrix phase to a retained gate}\label{pr:gate}

\begin{proof}[\proofanchor{proof:phase-app}Proof of Lemma~\ref{lem:phase-app} (the port and gate samplers)]
We verify the line-mass and sampler hypotheses of Lemma~\ref{lem:local}.
At a port every line has mass \(1/P\), so take \(D=P\). At a gate the prepared line determines both operand lines. For a product, these are the factor lines of a nonzero tensor. For \(c-s_1s_2\), project onto the tensor block first. For addition, project onto the two operand blocks. The operand lines on each side are independent and uniform, giving mass at most \(1/P^2\) and hence \(D=P^2\).

A uniform matrix meets the sampler hypothesis exactly. For a short product phase \(d=m\ge n^2\) and \(n\ge9\). Then \(\lfloor n^2/2\rfloor-3\ge40(n-1)/9\). Since \(P\le2q^{n-1}\), Lemma~\ref{lem:shear} bounds the mixed-image error by \(Cq^{-40(n-1)/9}\le CD^{-20/9}\). For the five-sample addition variant in \hyperref[claim:gate-costs]{Section~\ref*{sec:gate}}, \eqref{eq:repetition} in dimension \(2n\) bounds the error by \(Cq^{5(3-n)}\le Cq^{-40(n-1)/9}\le CD^{-20/9}\) when \(n\ge19\). Lemma~\ref{lem:local} now gives the claimed approximation in every case.
\end{proof}

\begin{proof}[\proofanchor{proof:retained}Proof of Lemma~\ref{lem:retained} (the answer given the wire encodings)]
We restrict Lemma~\ref{lem:phase-app} to encodings of the specified values. This proves \eqref{eq:retained} for a port. For a gate we then transfer the estimate through shortening.

\partitle{Restriction to fixed plaintexts}
On any fixed valid labeling the endpoint distribution for a port is the product of two prescribed orbits. For the long gate analysis use the product of the two prescribed input orbits and the \emph{uniform} long output orbit of the gate value. Relative to four independent vector arguments for a port, or the independent reference arguments for a gate, their densities are at most \((2q)^2\) and \((2q)^3\), respectively, by \eqref{eq:orbit-density}. On their support \(J=1\). Lemma~\ref{lem:phase-app} therefore gives \eqref{eq:retained} for the port and the analogous bound \(Cq^{3/2}P^{-2/9}\) for first-phase \(p-G\) on the specified long measure. Its balanced factors each have mean energy \(\nuc G\le C\) by Appendix~\ref{app:factors}.

\partitle{\proofanchor{proof:freshness}Freshness of the \hyperref[claim:shortening]{phase-two output}}
Write \(s_3\) for the prescribed gate output. The short pair satisfies \(l^\top r=U^\top V=s_3\), and the fallback when \(r=0\) preserves the zero product. For every fixed pair entering the final change of basis \(Q\), Lemma~\ref{lem:orbit} makes the short output uniform on \(\orb n{s_3}\). Hence fixing the operands and short output leaves the joint law of the long pair and \(\alpha\) unchanged.

\partitle{\proofanchor{proof:reverse}The \hyperref[claim:reverse]{reverse law for shortening}}
Let the long pair have the uniform orbit law just used in the \(L^2\) estimate. In this paragraph write \(s=s_3\), and define the regular event \(E=\{Z\ne0,\ r\ne0\}\) before the fallback assignment. Uniformity of the individual long shares gives
\[
         \epsilon_s=\Prb(E^c)\le q^{-k}+q^{1-n},
\]
since \(Z=0\) requires \(U_1=0\), and \(r=0\) requires \(V_0=0\). By exact short-output freshness this probability is unchanged on fixing any valid \(A,B\).

Given \(A,B,E\), the following is the exact reverse law of the ideal long pair and shortening coins. Draw \(Q\) uniformly subject to \(\alpha=(Q^{-1}A)_n\ne0\), put \((U_0,\alpha)=Q^{-1}A\), \((V_0,t)=Q^\top B\), and choose \(Z\) uniformly nonzero in \(\field^k\). Set \(U_1=\alpha Z\) and draw \(V_1\) uniformly on the hyperplane \(Z^\top V_1=t\). There are exactly \(q^{k-1}\) completions for every allowed \(Q,Z\). Each has the same weight in the forward experiment. The two long vectors are nonzero since \(U_1\ne0\) and \((V_0,t)\ne0\), and their pairing is \(s\). This proves the reverse law. For an unrestricted uniform \(Q\) the normalizing probability of its sole constraint is
\[
                        a_*=\frac{(q-1)q^{n-1}}{N}\ge\tfrac12,
\]
independent of the label, including at zero.

\partitle{The complete answer in \eqref{eq:retained}}
Fix the complete leaf \(z=(z_1,z_2)\) of the two phases. Let \(p\) be the first-phase table for \(z_1\), formed from the actual sampler. In the reverse law define \(\mathcal T H\) as the conditional expectation of \(H(x,U;y,V)\) multiplied by the phase-two answer indicators for \(z_2\). The latter are the functions specified following \(z_1\). On \(E\) the flag equals zero. Conditional on \(Q,Z\), the left indicator uses only \(x,A,U\) and the local coins and messages they specify, and the right indicator only \(y,B,V\) and its local data. Any additional rejection tape for sampling \(Q\) or \(\alpha\) can be averaged on the left conditional on the sampled values. The parties have erased the first-phase sampling tape and matrix description before the second phase.

For a fixed compatible tuple of short encodings, compare the actual probability of the joint answer with the calculation that draws the long pair from its uniform orbit. Let \(\eta\) be the forward \((2,2)\) error bounded in the \hyperref[proof:phase-app]{proof of Lemma~\ref*{lem:phase-app}}. This test includes one pair (Definition~\ref{def:mixed}), so at fixed operands the two long-pair laws are within \(\eta\). By exact freshness the same holds after fixing the short output. Both expectations use the same bounded integrand. It is the first-phase weight \(p\) times the probability of the second answer given the long shares and short output. Erasure and independent shortening coins justify this factorization. On long states unreachable in the actual phase use the zero extension of \(p\). Conditioning the uniform-orbit experiment on \(E\) then costs at most \(\epsilon_s\), by the test bound \eqref{eq:tv-test}. Thus
\[
 |\ell_z-\mathcal T p|\le \eta+\epsilon_s,\qquad
 \|\mathcal T(p-G)\|_{2,\mathrm{short},\boldsymbol s}
 \le (1-\epsilon_s)^{-1/2}\|p-G\|_{2,\mathrm{long},\boldsymbol s}.
\]
The second inequality is Jensen followed by averaging over the short orbit. This restores the long uniform-orbit experiment conditioned on \(E\). Working with the joint answer has kept rare first-phase leaves at their original weight.

Set \(G_z=\mathcal TG\). To give its product representation, sample \(Q\) without restriction and insert the left indicator \(\ind_{\{\alpha\ne0\}}/a_*\). Conditional on \(\omega=(Q,Z)\), each old term \(r_j(x,U)t_j(y,V)\), with the two answer indicators, has separate conditional expectations on the two sides. Split the weight \(\Prb(\omega)/a_*\) equally between the factors by assigning its square root to each. This gives a common representation on all short data. Its mean energies remain bounded. On the left, for uniform \(A\), the accepted inverse coordinates have \(U_0\) uniform and \(\alpha\) uniform nonzero. Averaging over \(Z\) makes \(U\) uniform conditional on \(U_1\ne0\). On the right, for uniform \(B\) at any fixed \(Q,Z\), the resulting \(V\) is uniform subject to \((V_0,Z^\top V_1)\ne0\). Both laws have density at most two relative to uniform on \(\Omega_m\), and \(a_*^{-1}\le2\). Applying Jensen to the individual factors proves the two mean-energy bounds. Finally \(\eta+\epsilon_s\le C q^{3/2}P^{-2/9}\) and \(1-\epsilon_s\) is bounded below. This proves the gate case of \eqref{eq:retained}.
\end{proof}

\begin{proof}[\proofanchor{proof:gate}Proof of Corollary~\ref{thm:gate}]
It remains to compare visible leakage for different inputs. Under the product reference measure for one wire, its encoding density is \(J_{n,s}/\pi_{n,s}\) by \eqref{eq:orbit-density}. This kernel has operator norm one by its common row and column mean and positivity. By \eqref{eq:orbit-mask} its difference from the constant-one kernel has norm at most
\begin{equation}
               \gamma=CqN^{-1/2}\le CqP^{-1/2}.               \label{eq:gamma}
\end{equation}
For a single gate the product of the three wire densities is a tensor product. Replacing its three factors successively by the constant kernel costs at most \(3\gamma\nuc{G_z}\le C\gamma\) in the mean of \(G_z\), by duality \eqref{eq:duality}, the nuclear norm estimate \eqref{eq:schur} and the tensor-product rule \eqref{eq:tensor-rule}. The resulting mean is common to all labels. Add the \(L^2\) error of Lemma~\ref{lem:retained} for each label, and sum over at most \(K_G\) complete answers. The triangle inequality gives the claimed transcript bound.
\end{proof}

\partitle{\proofanchor{proof:dimension}Dimension of the \hyperref[claim:gate-costs]{local gate preparations in Section~\ref*{sec:gate}}}
For fixed local preparations \(a,b\) of length \(m\), the table \(a_x^\top b_y\), over the \emph{base field}, has rank at most \(m\). For multiplication, choosing the left operands \(e_i,e_j\) and the right operands \(e_k,e_l\) exhibits an identity submatrix of order \(n^2\). For \(c-s_1s_2\), \(c\ne0\), the exact rank is \(n^2+1\). On either side the tensor coordinates span \(n^2\) independent functions, and their span excludes the constant one. With the second operand fixed, a linear function of the first cannot equal one at \(e_1,e_2,e_1+e_2\). Appending the constant gives full column rank on each side, with the nonsingular middle matrix \(\operatorname{diag}(-1,\ldots,-1,c)\). These are dimension bounds for local preparations followed by an inner product.

\partitle{\proofanchor{proof:implementation}Costs of the \hyperref[claim:gate-costs]{gate implementations}}
The messages counted in Section~\ref{sec:gate} are the long matrix description, the \(k\)-symbol tail, the flag and the \(n^2\) entries of \(Q\). For five short samples in dimension \(2n\), the long description costs \(5\cdot13n\) symbols. The tail has \(n+1\) coordinates, giving \(n^2+66n+1\) symbols and the flag for addition.

With a fixed public polynomial basis, a multiplication in the extension and its transpose for a fixed coefficient cost \(O(v^2)\) base-field operations by polynomial multiplication and reduction. Each of the three required powers of Frobenius can also be applied, or transposed, as a fixed linear map of that cost. The four shears thus cost \(O(m^2)\) field operations, including the action on covectors. A uniform short invertible matrix can be sampled by rejection, with fewer than four trials in expectation. Classical elimination and the applications of \(Q,Q^{-\top}\) cost expected \(O(n^3)\). This gives expected \(O(n^4)\) field operations for a product gate, \(O(n^3)\) for addition, and \(O(n^2)\) for a standalone short refresh. Field-operation counts use the public representations as parameters.

\subsection{Composition and choice of parameters}\label{pr:circuit}

Here is the quantitative form of Theorem~\ref{thm:circuit}. Write \(L=\lambda\) for the per-party operation allowance, or \(L=b\) for an equal allowance in every phase. With \(x=2^L P^{-1/H}\), the distance of the simulated and real views is at most
\begin{equation}
                 \min\{1,\ CSq^2(x^{c_R}+x^{c_G}+x^{c_W})\}.   \label{eq:circuit-main}
\end{equation}
The \hyperref[proof:circuit]{proof of this bound} uses the following parameters.
\[
\begin{array}{lrrrr}
\toprule
\text{Per-party allowance}&H&c_R&c_G&c_W\\
\midrule
\text{One budget per operation}&230&10&12&6\\
\text{The same budget in each phase}&305&14&14&8\\
\bottomrule
\end{array}
\]

\begin{proof}[\proofanchor{proof:circuit}Proof of Theorem~\ref{thm:circuit} and bound~\eqref{eq:circuit-main}]
\proofanchor{proof:canonical}We begin with the transcript identity \eqref{eq:canonical}.
The \hyperref[proof:freshness]{freshness calculation for gates}, Lemma~\ref{lem:orbit} for port refreshes, and the public sampler give each operation its prescribed output law for every preceding state. Factoring the joint law into this output law and the conditional answer probabilities gives \eqref{eq:canonical} in chronological order. At every valid tuple of endpoints the answer probabilities sum to one, including at a public sampler with a history-dependent input. Fix the adversary's tape throughout.

\partitle{Clipping the approximation in \eqref{eq:retained}}
For \(i\) of type \(R\) (port refresh) or \(G\) (gate), choose \(B_i\ge1\). In the factor representation from Lemma~\ref{lem:retained} zero the left factors on rows with energy above \(B_i\), and similarly the right factors on columns. Denote the resulting common table by \(\bar G_i\), and put \(e_i=\ell_i-\bar G_i\), suppressing the answer and history in the notation. Under any valid labeling the marginal on each side is precisely the uniform one used for the mean energies. Markov's inequality bounds the probability that a row or column is removed by \(C/B_i\). On those tuples \(|e_i|\le1\). On the remaining tuples \(|e_i|\le1+B_i\) and \(e_i=\ell_i-G_i\). Consequently, for \(t_i\ge2\), writing \(\varepsilon_i\) for a uniform upper bound on the corresponding \(L^{t_i}\) norm, we can take
\begin{equation}
 \varepsilon_i^{\,t_i}
    =(1+B_i)^{t_i-2}\delta_i^2+C/B_i .                \label{eq:clip}
\end{equation}
Take \(t_R=r,t_G=g\) with \(2/r+1/g\le1\). By the \hyperref[claim:wire-schedule]{port schedule in Section~\ref*{sec:circuit}}, a wire touches at most two ports and one gate, so the product-space H\"older bound \eqref{eq:finner} applies to any product of errors under independent wire encodings.

\partitle{Expanding the common expression}
Fix a formal transcript \(z\). Let \(\mathcal I\) be the gates and port refreshes. Let \(\mathcal V\) contain all touched wires except public sampler outputs, so it includes initial private wires. Write \(\mu_w=\law n{s_z(w)}\) and \(\Delta_w=\mu_w-\nu_n\). By \eqref{eq:gamma} the density of \(\Delta_w\) relative to \(\nu_n\) has operator norm at most \(\gamma\).

In \eqref{eq:canonical} use \(\bar G_i\) on \(\mathcal I\) and \(\nu_n\) on \(\mathcal V\). The resulting number \(C(z)\) is common to all private labelings with the same public choices, since the remaining factors and laws describe the public samplers. First expand \(\nu_n=\mu_w-\Delta_w\) on the eligible wires. Let \(W\) be the wires where we select \(-\Delta_w\), and \(N(W)\) the operations touching them. Keep \(\bar G_i\) on \(N(W)\). On the remaining gates and ports expand \(\bar G_i=\ell_i-e_i\), marking a set \(A\) of errors. This gives
\begin{equation}
 \begin{aligned}
 C(z)&=\sum_{W\subseteq\mathcal V}\ \sum_{A\subseteq\mathcal I\setminus N(W)}
              (-1)^{|W|+|A|} I_{A,W}(z),\\
 I_{A,W}(z)&=\int
       \prod_{i\in A}e_i\,
       \prod_{i\in N(W)}\bar G_i\,
       \prod_{i\notin A\cup N(W)}\ell_i
       \prod_{w\in W}d\Delta_w\prod_{w\notin W}d\mu_w .
 \end{aligned}                                                    \label{eq:marked-expansion}
\end{equation}
The empty choice gives \(I_{\varnothing,\varnothing}(z)=\Prb[Z=z]\). In every other term the wires used by a marked error still have their true laws, and only separate left and right factors meet a signed measure.

Fix the encodings outside \(W\). By the product-of-factors rule in Appendix~\ref{app:factors}, multiplying the representations on \(N(W)\) gives pointwise energies on both sides at most \((B_GB_R^2)^{|W|}\), since each marked wire touches at most one gate and two ports. The nuclear norm obeys the same bound by \eqref{eq:schur}. The product rule \eqref{eq:tensor-rule} bounds the operator norm of the tensor of signed densities by \(\gamma^{|W|}\). By \eqref{eq:duality}, integrating over \(W\) before taking absolute values costs at most \((\gamma B_GB_R^2)^{|W|}\).

Let \(K_R,K_G\) bound the answer alphabets of one-phase operations (including public sampling) and of gates, respectively. Enumerate answers of all operations incident either to the scope of a marked error or to a marked location. For one marked gate, enumerate the gate itself, the two refreshes that produced its input copies, the at most two next refreshes of those copies, and the first refresh of its output. At a marked port, its old copy has at most one neighboring gate and one other one-phase operation, and likewise its new copy. A marked private wire alone meets at most one gate and two one-phase operations. Allowing overcount at overlaps gives the following costs per marking.
\begin{equation}
 u_G=K_GK_R^5\varepsilon_G,\qquad
 u_R=K_G^2K_R^3\varepsilon_R,\qquad
 v=K_GK_R^2\,\gamma B_GB_R^2.                                \label{eq:marked-cost}
\end{equation}

To sum over adaptive transcripts, number the possible answers at each incoming history. Fix the answer numbers for the enumerated operations and bound their true probability factors by one. Every remaining true factor is disjoint from the wires of the marked errors. At a fixed transcript these wires have independent true encoding laws, so \eqref{eq:finner} bounds the mean product of absolute errors by \(\prod_{i\in A}\varepsilon_i\), uniformly over the history.

With the enumerated answer numbers fixed, the remaining true factors and wire laws have total mass at most one. Sum in reverse time, summing each operation's answer before integrating its output encoding. A true answer table sums to one for fixed valid endpoints. Its output law in turn has mass one for each label fixed by the preceding history. A fixed answer number at an enumerated operation has at most one continuation there. This argument allows later queries and public inputs to depend on earlier answers. Summing \(|I_{A,W}(z)|\) over \(z\) therefore costs at most the product of the marking costs in \eqref{eq:marked-cost}.

To bound \(\sum_z|\Prb[Z=z]-C(z)|\), sum the nonempty terms in \eqref{eq:marked-expansion}. There are at most \(S\) operations and \(3S\) touched locations. The sum is at most \((1+u_G)^S(1+u_R)^S(1+v)^{3S}-1\), which is \(O(S(u_G+u_R+3v))\) when \(S(u_G+u_R+3v)\le1\). Comparing two labelings through \(C(z)\) and using the unit bound on total variation otherwise gives
\begin{equation}
       \TV\le \min\{1,\ CS(u_G+u_R+v)\}.                      \label{eq:comp-general}
\end{equation}

\partitle{Leakage rates in \eqref{eq:circuit-main}}
For an operation budget \(L=\lambda\) we have \(K_R,K_G\le2^{2L}\). With a phase budget \(L=b\) we have \(K_R\le2^{2L}, K_G\le2^{4L}\). Reading off \eqref{eq:marked-cost}, the alphabet costs of a port error, gate error and wire difference are \(2^{c_RL},2^{c_GL},2^{c_WL}\), with the triples displayed after \eqref{eq:circuit-main}. Take \(r>2,\ g=r/(r-2)\). For a target denominator \(H\) set
\[
                      B_R=P^{c_R r/H},\qquad B_G=P^{c_G g/H}.
\]
Equations \eqref{eq:clip} and \eqref{eq:gamma} give \(\varepsilon_i\le Cq^2P^{-c_i/H}\) and \(\gamma B_G B_R^2\le CqP^{-c_W/H}\) provided
\begin{equation}
 \max\left\{\tfrac92 c_Rr(r-1),\ \tfrac94 c_Gg(g-1),\
                 2(c_W+2c_Rr+c_Gg)\right\}\le H .
 \label{eq:exponents}
\end{equation}
The first two conditions make \(\delta_i^2 B_i^{t_i-2}\) at most a constant times \(q^3/B_i\) for \(i=R,G\). To choose the exponents, write \(r=2+u,\ g=1+2/u\). Balancing the first two terms of \eqref{eq:exponents} gives \(u^2(1+u)=c_G/c_R\). Nearby rational choices verify all three conditions. For the operation budget take \((r,g)=(76/27,38/11)\), giving maximum \(18620/81<230\). For the phase budget take \((r,g)=(113/41,113/31)\), giving \(512568/1681<305\). This proves \eqref{eq:circuit-main}. Since \(\log_2 P=(n-1)\log_2q+O(1)\), it gives exponential decay below the two rates of Theorem~\ref{thm:circuit}, uniformly in \(q\).

\partitle{\proofanchor{proof:simulation}Analysis of the \hyperref[claim:simulator]{simulator in Section~\ref*{sec:circuit}}}
Run the compiled protocol on zero private input or zero initial state, retaining the dummy encoded state across invocations. At the end of an invocation return the response of the ideal circuit on the actual public input. For a fixed adversary tape and a fixed true initial state, the public inputs and authorized responses are determined by the formal answer prefix in both experiments. Both induced labelings are valid and agree on the public sampled locations. Equation~\eqref{eq:comp-general} compares their leakage, and adjoining the same functions of the transcript preserves the distance. A chronological unrolling counts all uses of persistent wires in their port chains and applies the same comparison once to the entire interaction.

To sample an initial or public encoding, draw the left vector uniformly nonzero, select a nonzero coordinate and solve the corresponding linear equation for the missing coordinate of the right vector after sampling its other coordinates uniformly. Reject the right vector if it is zero. This has the required uniform law and constant expected trials. The same holds for invertible matrix sampling. The \hyperref[proof:implementation]{work bounds in Section~\ref*{pr:gate}} cover the internal operations under standard polynomial-basis field representations. The simulator also samples one encoding per initial private input or state element, reads the public inputs, and forwards the authorized outputs. Including these input and output lengths gives the expected work bound of Theorem~\ref{thm:circuit}. In the basic one-query-per-party model each source gate and its two input-port refreshes entail eight local answer evaluations, and each public sampler entails two. Extra queries in the interleaved model are charged to their actual evaluation cost. This completes the proof.
\end{proof}

\clearpage
\phantomsection

\clearpage
\appendix
\section{Spectral tools on finite probability spaces}\label{sec:tools}

This appendix gives our SVD conventions, the norm bounds for separate left and right factors, and the finite-field and probability estimates used in the proofs. For standard matrix conventions see \cite[Sec.~1.1.3]{Watrous}.

\subsection{From a table to its singular functions}\label{app:svd}

Give the nonempty finite sets \(\mathcal X,\mathcal Y\) uniform measures, with inner product \(\langle f,g\rangle=\E(fg)\). A table acts on a function \(g\) on \(\mathcal Y\) by \((T_Hg)(x)=\E_y H(x,y)g(y)\), the average of row \(x\) against \(g\). In normalized point-indicator bases its matrix is \(H/\sqrt{|\mathcal X||\mathcal Y|}\). We take the singular values of this map. Its SVD reads
\[
 H(x,y)=\sum_{j=1}^{r}\sigma_j e_j(x)h_j(y),
 \qquad \sigma_1\ge\cdots\ge\sigma_r>0 .
\]
Here \(e_j\) and \(h_j\) are functions of the row and column indices respectively. Each family is orthonormal for the corresponding uniform measure.

For example, the indicator of a rectangle \(A\times B\) with positive marginal masses \(a,b\) has one singular value \(\sqrt{ab}\). Its singular functions are \(\ind_A/\sqrt a\) and \(\ind_B/\sqrt b\). Thus the constant-one table has singular value one in every dimension. More generally, at a fixed row \(x\), orthonormality of the \(h_j\)'s gives
\begin{equation}
          \E_y|H(x,y)|^2=\sum_j\sigma_j^2|e_j(x)|^2 .
          \label{eq:row-pythagoras}
\end{equation}
If this row has mean square at most \(R\), then at any cutoff \(\tau>0\) we have
\begin{equation}
 \sum_{\sigma_j\ge\tau}\sigma_j e_j(x)^2\le R/\tau,\qquad
 \sum_{\sigma_j\ge\tau} e_j(x)^2\le R/\tau^2 .                    \label{eq:truncation-energies}
\end{equation}
Indeed, on the kept terms \(\sigma_j^2\ge\tau\sigma_j\ge\tau^2\). All three formulas have column counterparts. They give the storage cutoff bound \eqref{eq:storage-trunc} and the gate energies \eqref{eq:phase-energies}.

\subsection{Which norm controls which average?}\label{app:norms}

The operator, Hilbert--Schmidt, fourth Schatten and nuclear norms are, respectively,
\[
 \op H=\max_j\sigma_j,\qquad
 \hs H=(\textstyle\sum_j\sigma_j^2)^{1/2},\qquad
 \siv H=(\textstyle\sum_j\sigma_j^4)^{1/4},\qquad
 \nuc H=\textstyle\sum_j\sigma_j .
\]
All four norms vanish on the zero table, and \(\hs H^2=\E|H|^2\). We also use \(\ellone H=\E|H|\), \(\supn H=\max|H|\) and \(\|H\|_{L^p}=(\E|H|^p)^{1/p}\). In particular, \(L^4\) averages fourth powers of entries, while \(S_4\) uses singular values. Writing \(\langle H,G\rangle=\E(HG)\), we have
\begin{align}
 |\langle H,G\rangle|&\le\op H\,\nuc G,                         \label{eq:duality}\\
 \op H&\le\siv H\le\hs H,\qquad \siv H^4\le\op H^2\hs H^2.     \label{eq:schatten}
\end{align}
Expand \(G\) in singular functions to obtain \eqref{eq:duality}. For \(G(x,y)=f(x)g(y)\), its nuclear norm is \(\|f\|_2\|g\|_2\). Inequalities \eqref{eq:schatten} follow term by term. Truncation at \(\tau\) leaves operator error at most \(\tau\). To control the entrywise error after one more refresh we use Lemma~\ref{lem:smooth}, or \eqref{eq:general-smoothing} for a short sampler.

Choose independent uniform \(x,x'\in\mathcal X\) and \(y,y'\in\mathcal Y\). The four-corner formula is
\begin{equation}
 \begin{aligned}
 \cyc(H)=\siv H^4
  &=\E\,H(x,y)H(x',y)H(x',y')H(x,y')\\
  &=\E_{x,x'}\bigl(\E_y H(x,y)H(x',y)\bigr)^2 .
 \end{aligned}                                                     \label{eq:cycle}
\end{equation}
To verify \eqref{eq:cycle}, observe that \(T_HT_H^\top\) has kernel \(\E_yH(x,y)H(x',y)\) and singular values \(\sigma_j^2\). The resulting mean square controls smoothing in \eqref{eq:general-smoothing} and the gate estimate \eqref{eq:matrix4}.

\subsection{Products of separate factors}\label{app:factors}

For a representation \(G(x,y)=\sum_j r_j(x)t_j(y)\), its pointwise energies are \(\Phi(x)=\sum_j r_j(x)^2\) and \(\Psi(y)=\sum_j t_j(y)^2\). The balanced factors \(r_j=\sqrt{\sigma_j}e_j,\ t_j=\sqrt{\sigma_j}h_j\) have mean energy \(\nuc G\) on each side. For any representation, and for kernels on independent coordinates, we have
\begin{align}
 \op{K\odot G}&\le e\op K\quad\text{if }\Phi,\Psi\le e,
 \qquad \nuc G\le\sqrt{\E\Phi\,\E\Psi},                         \label{eq:schur}\\
 \op{K_1\otimes K_2}&=\op{K_1}\op{K_2}.                         \label{eq:tensor-rule}
\end{align}
Here \((K\odot G)(x,y)=K(x,y)G(x,y)\). For the first bound in \eqref{eq:schur}, test on two unit functions \(a,b\), expand \(G\) and apply Cauchy--Schwarz to \(\sum_j\|ar_j\|_2\|bt_j\|_2\). For the second, sum the rank-one nuclear norms \(\|r_j\|_2\|t_j\|_2\) and apply Cauchy--Schwarz. Identity~\eqref{eq:tensor-rule} follows because the singular values of a tensor product are the pairwise products.

On overlapping coordinates, multiplying two separate representations gives factors \(r_jr'_k,t_jt'_k\), so their pointwise energies multiply. The circuit proof uses this fact and \eqref{eq:tensor-rule} to bound the cost of a marked wire in \eqref{eq:marked-cost}.

\subsection{Hyperplane incidence and probability}\label{app:incidence}

\partitle{A hyperplane bound}
For \(t\in\field\), \(r\in\field^m\setminus\{0\}\) and \(z\in\field^m\), put \(C_t(r,z)=q\ind_{\{r^\top z=t\}}-1\). Let the independent indices \(R,Z\) have largest atom masses \(a,b\), respectively. The operator \(f\mapsto\bigl(r\mapsto\E_Z[C_t(r,Z)f(Z)]\bigr)\) between their probability \(L^2\) spaces satisfies
\begin{equation}
                    \op{C_t}\le(q-1)q^{m/2}\sqrt{ab}.              \label{eq:hyperplane}
\end{equation}
To prove \eqref{eq:hyperplane}, first use counting measure. For uniform \(z\) the form \(r^\top z\) is uniform, so each row has mean zero and squared length \((q-1)q^m\). For nonproportional \(r,r'\) the forms are independent, hence the rows are orthogonal. Each block of the Gram matrix \(C_tC_t^\top\) corresponds to a line of \(q-1\) indices. It is positive semidefinite, so its largest eigenvalue is at most its trace \((q-1)^2q^m\). Taking square roots gives the counting-measure bound, also at \(t=0\).

In normalized point-indicator bases, each row and column is scaled by the square root of its mass, giving \eqref{eq:hyperplane}. Indices may be functions of larger local arguments since conditional averaging is an \(L^2\) contraction. See the storage mask \eqref{eq:orbit-mask} and the gate mask \eqref{eq:phase-mask}.

\partitle{Probability inequalities}
For probability laws on a finite set put \(\eta=\TV(\mu,\nu)=\tfrac12\sum_x|\mu(x)-\nu(x)|\). The positive and negative parts of \(\mu-\nu\) each have mass \(\eta\). Thus for any real test \(f\),
\begin{equation}
             |\E_\mu f-\E_\nu f|\le\eta(\max f-\min f).          \label{eq:tv-test}
\end{equation}
This costs \(\eta\) for a test in \([0,1]\) and \(2\eta\) for a test in \([-1,1]\), as in \eqref{eq:history-weights} and \eqref{eq:general-smoothing}, respectively.

For independent coordinates \(X_w\) and functions \(f_i\) with scopes \(D_i\), the product-space H\"older inequality \cite{Finner} gives
\begin{equation}
   \E\prod_i|f_i(X_{D_i})|\ \le\ \prod_i\|f_i\|_{p_i}
   \quad\text{if}\quad
   \sum_{i:\,w\in D_i}\frac1{p_i}\le1\quad\text{for every }w,
   \qquad p_i\ge1.                                                \label{eq:finner}
\end{equation}
To prove \eqref{eq:finner}, integrate one coordinate at a time. Ordinary H\"older applies to its incident factors, with the constant one filling any unused reciprocal exponent. The resulting factors are partial \(L^{p_i}\) norms. In the \hyperref[proof:circuit]{proof of Theorem~\ref*{thm:circuit}}, the coordinates are wires and \(D_i\) is the set of wires incident to operation \(i\).
\end{document}